\documentclass[aps,pre,preprint,onecolumn,citeautoscript,superscriptaddress,
eqsecnum]{revtex4-1}  
\usepackage{amsmath,amssymb,bm} 
\usepackage{graphicx}  
\usepackage[tight]{subfigure}    
\usepackage{color} 
\usepackage[papersize={8.5in,11in}]{geometry}
\usepackage[colorlinks=true]{hyperref}  
\hypersetup{
    bookmarks=true,         % show bookmarks bar?
    unicode=false,          % non-Latin characters 
    pdftoolbar=true,        % show Acrobat
    pdfmenubar=true,        % show Acrobat 
    pdffitwindow=false,     % window fit to page when opened
    pdfstartview={FitH},    % fits the width of the page to the window
    pdftitle={Spin density wave order, topological order, and Fermi surface reconstruction},    % title
    pdfauthor={Subir Sachdev, Erez Berg, Shubhayu Chatterjee, and Yoni Schattner},     % author
    pdfsubject={},   % subject of the document
    pdfcreator={Creator},   % creator of the document
    pdfproducer={Producer}, % producer of the document
    pdfkeywords={} {} {}, % list of keywords
    pdfnewwindow=true,      % links in new window
    colorlinks=true,       % false: boxed links; true: colored links
    linkcolor=magenta, %red,          % color of internal links (change box color with linkbordercolor)
    citecolor=blue,        % color of links to bibliography
    filecolor=magenta,      % color of file links
    urlcolor=blue           % color of external links
} 

\usepackage{amsfonts}
\usepackage{upgreek}
\usepackage{slashed}
\usepackage{latexsym}
\usepackage{braket}

\usepackage{todonotes}

\newcommand{\beq}{\begin{equation}}
\newcommand{\eeq}{\end{equation}}
\def\bea{\begin{eqnarray}}
\def\eea{\end{eqnarray}}
\def \ua{\uparrow}
\def \da{\downarrow}
\newcommand{\nn}{\nonumber \\}

\newcommand{\br}{{\bm r}}

\newcommand{\bk}{{\bm k}}
\newcommand{\bp}{{\bm p}}

\newcommand{\bq}{{\bm q}}
\newcommand{\bK}{{\bm K}}

\begin{document}

\preprint{\href{http://arXiv.org/abs/1606.07813}{arXiv:1606.07813}}
\title{Spin density wave order, topological order,\\ and Fermi surface reconstruction}

\author{Subir Sachdev}

\affiliation{Department of Physics, Harvard University, Cambridge MA 02138, USA}
\affiliation{Perimeter Institute for Theoretical Physics, Waterloo, Ontario, Canada N2L 2Y5}

\author{Erez Berg}
\affiliation{Department of Condensed Matter Physics, The Weizmann Institute of Science, Rehovot, 76100, Israel}

\author{Shubhayu Chatterjee}

\affiliation{Department of Physics, Harvard University, Cambridge MA 02138, USA}

\author{Yoni Schattner}
\affiliation{Department of Condensed Matter Physics, The Weizmann Institute of Science, Rehovot, 76100, Israel}

\date{\today
\\
\vspace{0.4in}}

\begin{abstract}%
In the conventional theory of density wave ordering in metals, the onset of spin density wave (SDW)
order co-incides with the reconstruction of the Fermi surfaces into small `pockets'. We present models
which display this transition, while also displaying an alternative route between these phases 
via an intermediate phase with topological order, 
no broken symmetry, and pocket Fermi surfaces. The models involve coupling emergent 
gauge fields to a fractionalized SDW order, but retain the canonical electron operator in the underlying Hamiltonian.
We establish an intimate connection between the suppression of certain defects in the SDW order, and the presence
of Fermi surface sizes distinct from the Luttinger value in Fermi liquids. We discuss the relevance of such models
to the physics of the hole-doped cuprates near optimal doping.
\end{abstract}

%\subjectindex{spin liquids,  quantum criticality, topological quantum field theory, Higgs transition}

\maketitle

\section{Introduction}
\label{sec:intro}

A number of recent experiments \cite{Marel13,YHe13,Fujita14a,LTCP15} 
have highlighted a remarkable transformation in the electronic state of
the hole-doped cuprates at a hole density around $p = p_c \approx 0.19$: many electronic properties change from those
characteristic of a Fermi gas of charge $+e$ carriers of density $p$ for $p<p_c$, to those of a Fermi gas of 
charge $+e$ carriers of density $1+p$ for $p>p_c$. 
As the density of holes is conventionally
measured relative to those of the insulator at unit density, a conventional Fermi liquid is required by the Luttinger theorem
to have a Fermi surface of size $1+p$, as found for $p > p_c$. 

Starting from the Fermi liquid with a Fermi surface of size $1+p$, there are two reasonable routes to 
a Fermi surface reconstruction of size $p$ that could apply to the cuprates:\\
({\em i\/}) The conventional route involves the onset of spin density wave (SDW) order
(other density wave orders have also been suggested \cite{SC03}), which reconstructs the ``large'' Fermi surface
to pocket Fermi surfaces. This route appears appropriate for the electron-doped cuprates, where antiferromagnetic order
is observed \cite{Greven07} not too far from the critical electron doping.\\
({\em ii\/}) The more `exotic' route relies on the development of topological quantum order in the metallic state, 
which has been linked to changes in the Fermi surface size \cite{TSSSMV03,TSMVSS04,APAV04}. 
This is an attractive and exciting possibility for the hole-doped
cuprates, given the absence in observations so far 
of significant correlations in any order parameter which breaks translational symmetry near $p=p_c$.

The purpose of this paper is to present the simplest models in which the 
existence of the three metallic phases mentioned above (the Fermi liquid, the Fermi liquid with SDW order,
and the metal with small Fermi surfaces and topological order)
can be reliably established. We wish to describe models which can serve as convenient starting points for analyzing 
the quantum phase transitions between these metals, and which do not have extraneous exotic phases which are ultimately
unstable to confinement. 

The models described below are closely connected to previous work \cite{SS09,DCSS15b,DCSS15,SSDC16} 
using a SU(2) gauge theory and a Higgs field
to represent local antiferromagnetic correlations in a metal. These previous works, along with related works using a Schwinger boson
formulation \cite{RKK07,RKK08} or a quantum dimer model \cite{Punk15}, 
show that the phases we obtain below are allowed ground states of a single-band Hubbard model.
However, in the interests of simplicity and of 
keeping this paper
self-contained, we will not introduce the
models using these prior connections. Instead,  we will emphasize the relationship of our models to those
using a conventional Landau-Ginzburg-Wilson (LGW) order parameter framework for the onset of spin
density wave order in metals; as emphasized by Hertz \cite{Hertz76}, the general rules of LGW theories apply
to such quantum phase transitions in metals, and the main role of the Fermi surface is to damp with dynamic order parameter fluctuations. 
Developing our model by deforming the LGW theory will clearly expose the intimate link between topological
defects in the SDW order, and the possibility of metallic states which have Fermi surface sizes distinct from the
Luttinger value. 

As the LGW-Hertz theory is an expansion around weak coupling, our analysis below can be viewed
as providing the minimal ingredients necessary to put strong coupling Mott-Hubbard physics back into the LGW-Hertz model.
Indeed in the limit of $p=0$, our model will have a Mott insulator with $\mathbb{Z}_2$ topological order \cite{NRSS91,XGW91} (and other topological
orders in related models), in addition to the `Slater' insulator with N\'eel order present in the LGW-Hertz model.

\subsection{Easy-plane model}
\label{sec:easy}

The most transparent introduction to our models is obtained by focusing on the case in which the spin density
wave order parameter is restricted to lie in the $x$-$y$ plane in spin space. Such a restriction can only arise
from spin-orbit couplings, which are known to be rather weak in the cuprates. Nevertheless, we will describe this
case first because of its simplicity.

\subsubsection{LGW-Hertz theory}

The LGW-Hertz theory for the onset of SDW order can be described by the following Hamiltonian
\beq
H_{\rm sdw} = H_c + H_\theta + H_Y,
\label{Hsdw}
\eeq
where $H_c$ describes electrons (of density $(1-p)$) hopping on the sites of a square lattice
\beq
H_c = - \sum_{i,j} \left(t_{ij} + \mu \delta_{ij} \right) c_{i \alpha}^\dagger c_{j \alpha} \label{Hc}
\eeq
with $c_{i \alpha}$ the electron annihilation operator on site $i$ with spin $\alpha = \uparrow, \downarrow$.
We represent the SDW order by a lattice XY rotor model, described by an angle $\theta_i$,
and its canonically conjugate number operator $N_i$, obeying
\beq
H_\theta = -  \sum_{i< j} J_{ij} \cos (\theta_i - \theta_j ) + 4\Delta \sum_i N_i^2 \quad ; \quad  [ \theta_i , N_j ] = i \delta_{ij},
\eeq
where $J_{ij}$ positive exchange constants, and $\Delta$ is proportional to the bare spin-wave gap (the 4 is for future convenience).
A term linear in $N_i$ is also allowed in $H_\theta$, but we ignore it for simplicity; such a linear term will not be allowed
when we consider models with SU(2) symmetry in Section~\ref{sec:su2}.
Finally, there is a `Yukawa' coupling between the XY order parameter, $e^{i \theta}$, and the fermions
\beq
H_Y = - \lambda \sum_i \eta_i \left[ e^{- i \theta_i} c_{i \uparrow}^\dagger c_{i \downarrow}
+ e^{ i \theta_i} c_{i \downarrow}^\dagger c_{i \uparrow} \right], \label{HY0}
\eeq
where 
\beq
\eta_i \equiv (-1)^{x_i + y_i}
\eeq
is the staggering factor representing the opposite spin orientations on the two sublattices. Note that the Yukawa coupling,
and the remaining Hamiltonian, commute with the total
spin along the $z$ direction
\beq
S_z = \sum_i \left( N_i + \frac{1}{2} c_{i \uparrow}^\dagger c_{i \uparrow} - \frac{1}{2} c_{i \downarrow}^\dagger c_{i \downarrow} \right).
\label{Szdef}
\eeq

\begin{figure}
\begin{center}
\includegraphics[height=10cm]{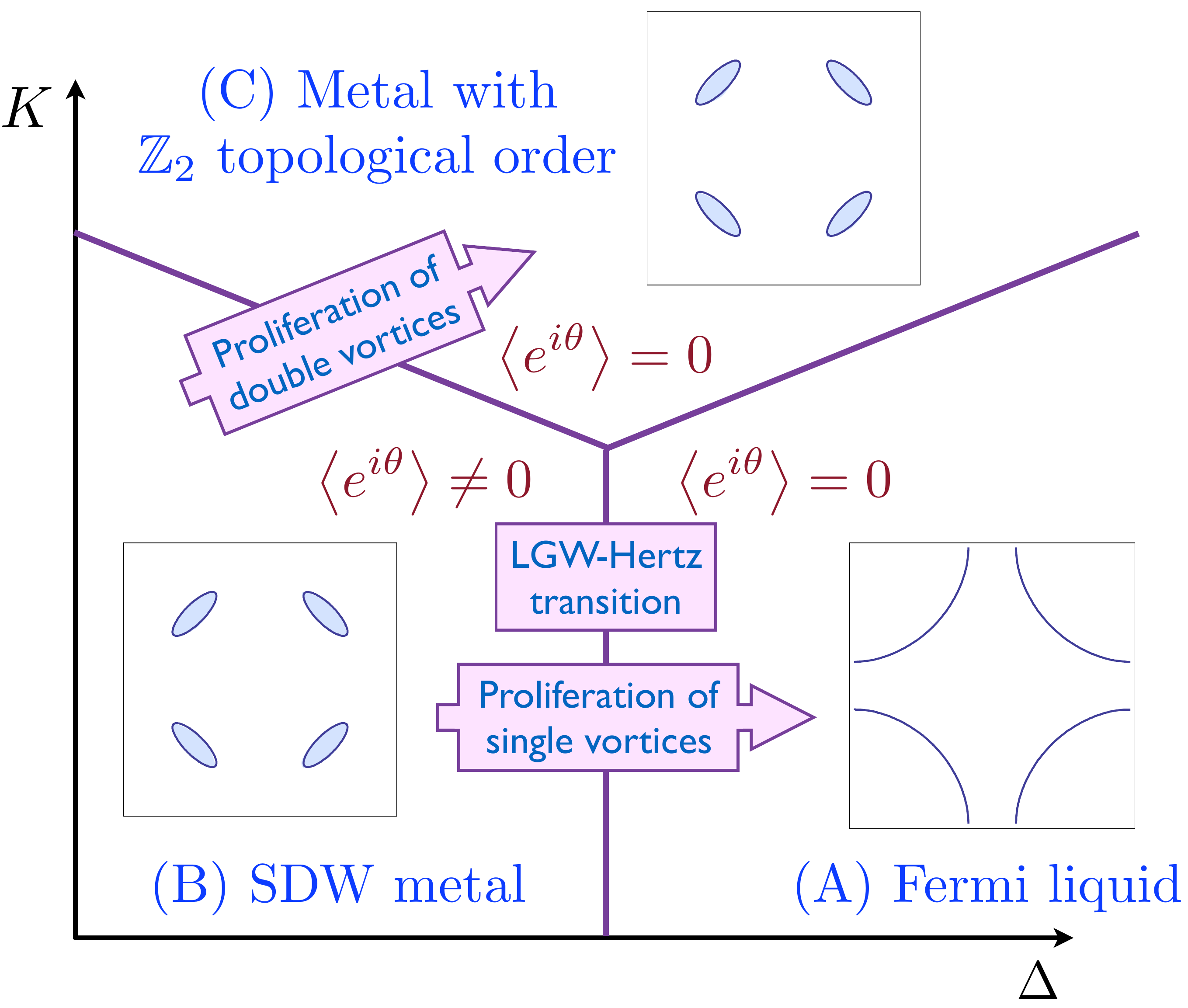}
\end{center}
\caption{Schematic, minimal, phase diagram of the easy-plane Hamiltonian $\mathcal{H}_1$ in Eq.~(\ref{H1}). 
The vortices are the usual defects in the XY SDW order $e^{i \theta}$.
The Fermi surfaces are shown in the first Brillouin zone: those in A and B are of electrons, while
those in C can be either of electrons or chargons. In phase C, the single vortices in the SDW order are gapped excitations,
identified as the visons of the $\mathbb{Z}_2$ topological order. The sketched Fermi surfaces are for hole-doping with the
cuprate band structure: in phases B and C only hole pockets are shown, but electron pockets will appear near the boundaries
to phase A. We propose that the SU(2) spin rotation invariant analogs of phase C (discussed in Section~\ref{sec:su2}) 
describe the pseudogap
state in the hole-doped cuprates. Other less-correlated superconductors (such as the pnictides) are proposed to bypass 
phase C and evolve directly from phase B to A.}
\label{fig:xy}
\end{figure}

The Hamiltonian $H_{\rm sdw}$ displays the two conventional metallic phases noted above.
These phases can be conveniently accessed by tuning the value of $\Delta/J$, where $J$ is the nearest-neighbor
exchange. For large $\Delta/J$, the correlations of $e^{i \theta}$ are short-ranged, and we obtain the Fermi liquid with
a large Fermi surface controlled mainly by $H_c$; we can account for $H_Y$ perturbatively in $\lambda$, 
and the large Fermi surface leads to damping in the order parameter correlation functions.
On the other hand, for small $\Delta/J$, we expect long-range XY order with $\left\langle e^{i \theta} \right\rangle \neq 0$;
now $H_Y$ has a stronger effect, and in the presence of the XY condensate the fermion dispersion is modified,
leading to a reconstruction of the Fermi surface into small pockets. These phases of $H_{\rm sdw}$ are illustrated
by the small $K$ regime of Fig.~\ref{fig:xy}, where they are labelled A and B respectively (the parameter $K$ will be introduced below).
The phase transition between these two phases has been extensively studied 
\cite{Millis93,abanov00,ACS03,MMSS10b,LeeStrack13,SurLee15,PatelStrack15,Strack16}
since the original work by Hertz \cite{Hertz76}, including by recent sign-problem-free quantum Monte Carlo 
simulations \cite{BMS12,SGTB15,LWYL15}.
Note that in such a phase transition, two important physical changes happen at the same point in the phase diagram:
the appearance of long-range XY order, and the reconstruction of the Fermi surface.

It is also interesting to consider the $p=0$ limit of phases A and B. The large Fermi surface has size $1+p$ and so phase A
is not sensitive to $p$ approaching 0: it remains a Fermi liquid. On the other hand, in phase B, the hole pockets disappear at $p=0$,
so phase B is an insulator. This insulating behavior is a direct consequence of the presence of strong long-range XY order, 
and so B should be considered a Slater insulator at $p=0$. We note that in between the Slater insulator and the large Fermi surface 
Fermi liquid, there is a metal with hole and electron pockets, and this is not shown in Fig.~\ref{fig:xy}.

\subsubsection{Fractionalizing the order parameter}
\label{sec:fracorder}

We now ask if it is possible, at non-zero $p$, 
to realize a situation in which XY long-range order and Fermi surface reconstructions happen
at distinct points of the phase diagram. If so, we will obtain an intermediate phase with small Fermi pockets
but no long-range XY order. (At $p=0$, such a phase would be an insulator without long-range XY order, and so would be a
Mott insulator.)
To obtain such a phase, we use the idea of transforming to a `rotating reference frame'
determined by the local orientation of the XY order \cite{SS88,SS89,SS09}. In particular, by a rotation about the 
$z$ axis in spin space, let us define the canonical fermion operators
\beq
\psi_+ = e^{i \theta/2} c_\uparrow \quad, \quad \psi_- = e^{-i \theta/2} c_\downarrow. \label{psidef}
\eeq
Then the Yukawa coupling, $H_Y$, takes a simple form independent of the orientation of the XY order \cite{SS09}:
\beq
H_Y = - \lambda \sum_i \eta_i \left[ \psi_{i+}^\dagger \psi_{i-} + \psi_{i-}^\dagger \psi_{i+} \right]. \label{Hsdw2}
\eeq
In other words, the $\psi_\pm$ fermions
move in the presence of a spacetime-independent XY order, even though the actual orientation of the XY order rotates
from point to point. Moreover, from the electron hopping term in $H_c$, we can obtain an effective hopping
$Z_{ij} t_{ij} ( \psi_{i+}^\dagger \psi_{j+} + \psi_{i-}^\dagger \psi_{j-})$ where $Z_{ij} = \langle e^{\pm i (\theta_i - \theta_j)/2} \rangle$ is a renormalization factor of order unity (computed later in Eq.~(\ref{Zh})). So it appears 
we can realize a situation in which the $\psi_{\pm}$ fermions are approximately free, and their observation
of constant XY order implies that they will
form small pocket Fermi surfaces (or be fully gapped at $p=0$). 
From (\ref{Szdef}), it can be verified that the $\psi_{\pm}$ fermions
have $S_z = 0$, and so these are spinless fermions which carry only the charge of the electron: we will refer to them
as `chargons' in the remaining discussion. A metallic phase with chargon Fermi surfaces was called an `algebraic charge liquid'
(ACL) in Ref.~\onlinecite{RKK08}. 

However, further thought based upon the structure of (\ref{psidef}) shows that there is a crucial obstacle to 
realizing nearly-free $\psi_\pm$ fermions in a regime where $\left\langle e^{i \theta} \right\rangle = 0$.
The phase with no XY order described by $H_{\rm sdw}$ has proliferating $2\pi$ vortices in the SDW order, 
and the half-angle transformation in 
Eq.~(\ref{psidef}) shows that $\psi_\pm$ are not single-valued around such vortices. So the $\psi_\pm$ fermions must be
confined at the same point where the XY order disappears. In other words, we are back to the conventional scenario in which
Fermi surface reconstruction and XY ordering co-incide.

But the above argument also suggests a route around such an obstacle:
the $\psi_\pm$ fermions are single-valued around {\em doubled\/} $4 \pi$ vortices, and so we need the disappearance of 
XY order to be associated with the proliferation of doubled vortices. 

There is a simple route to the loss of XY order by doubled vortices that has been much studied in the literature 
\cite{TSMPAF00,SSS02,KPSS02,KPSS02a}:
it involves coupling the square-root of the XY order, the `spinon' field $e^{i \theta/2}$, to a $\mathbb{Z}_2$ gauge field.
A microscopic justification for fractionalizing the order parameter in this manner can be obtained from the Schwinger
boson theory of frustrated antiferromagnets \cite{NRSS91,SSNR91,CSS94,SS09}: $e^{i \theta/2}$ is essentially the staggered Schwinger boson operator.
The model we wish to study is obtained by replacing $H_\theta$ in $H_{\rm sdw}$ by the model studied in 
Refs.~\onlinecite{TSMPAF00,SSS02}. 
In this manner, we obtain the Hamiltonian (written out completely because of our focus on it in this paper)
\bea
\mathcal{H}_1 &=& H_c + H_{\theta,\mathbb{Z}_2} + H_Y \nn
H_c &=& - \sum_{i,j} \left(t_{ij} + \mu \delta_{ij} \right) c_{i \alpha}^\dagger c_{j \alpha} \nn
H_Y &=& - \lambda \sum_i \eta_i \left[ e^{- i \theta_i} c_{i \uparrow}^\dagger c_{i \downarrow}
+ e^{ i \theta_i} c_{i \downarrow}^\dagger c_{i \uparrow} \right] \nn
H_{\theta,\mathbb{Z}_2} &=& -  \sum_{i< j} J_{ij} \mu^{z}_{ij} \cos \left((\theta_i - \theta_j )/2\right) + 4\Delta \sum_i N_i^2 
- g \sum_{\langle ij \rangle} \mu^x_{ij} - K  \sum_{\square} \left[\prod_{\square} \mu^z_{ij}\right],
\label{H1}
\eea
where $\mu^{x,z}$ are Pauli matrices on the links of the square lattice representing the $\mathbb{Z}_2$ gauge field. 
The forms of $H_c$ and $H_Y$ are the same as those in the LGW-Hertz theory, and only the action for the spin density 
wave order has been modified by terms that are effectively multi-spin exchange interactions.
(It will become clear from our discussion later that at $p=0$ and small $\Delta$, 
$\mathcal{H}_1$ reduces to the model studied in Ref.~\onlinecite{KPSS02a}.)
The Hamiltonian $\mathcal{H}_1$ is invariant under the $\mathbb{Z}_2$ gauge transformation
\beq
e^{i \theta_i/2} \rightarrow s_i \, e^{i \theta_i/2} \quad , \quad \mu^z_{ij} \rightarrow s_i \, \mu^z_{ij} \, s_j,
\eeq
where $s_i = \pm 1$ is an arbitrary function of $i$, and the other operators remain invariant. Associated with this gauge invariance
is the existence of an extensive number of conserved charges, $\hat{G}_i$, which commute with $\mathcal{H}_1$
and obey $\hat{G}_i^2 = 1$; we restrict
our attention to the gauge-invariant sector of the Hilbert space in which all the $\hat{G}_i = 1$:
\beq
\hat{G}_i \equiv e^{2 i \pi \hat{N}_i} \prod_{j \in {\rm n.n.}(i)} \mu^x_{ij} = 1, \label{Gi}
\eeq
where $j$ extends over the nearest-neighbors of $i$.

The main term driving the appearance of exotic phases in $\mathcal{H}_1$ is the $K$ term, which penalizes configurations
with non-zero $\mathbb{Z}_2$ gauge flux. For small $K$, we can trace over the $\mathbb{Z}_2$ gauge field in powers
of $K$, and then $\mathcal{H}_1$ only has terms with the same structure as those in $H_{\rm sdw}$. 
However, at large $K$, the suppression of $\mathbb{Z}_2$ gauge flux implies that single vortices (but not double vortices) 
in $e^{i \theta}$
become very expensive: the coupling $J_{ij}$ ties $\mathbb{Z}_2$ gauge flux to a $2 \pi$ vortex in $e^{i \theta}$
because of the branch cut in $e^{i \theta/2}$ around such a vortex.
Hence, upon increasing $\Delta$ at large $K$, we obtain the needed transition to a phase without XY order by
the proliferation of double vortices.  The resulting $\mathbb{Z}_2$ topologically ordered phase supports gapped 
deconfined ``spinon'' 
excitations that carry a half integer value of $N_i$, and gapped vison excitations 
which are the remnants of the single vortices in the SDW
ordered phase.
%In order to get the desired physical properties of this phase, we supplement the Hamiltonian (\ref{H1}) by the set of constraints
%\bea
%e^{2\pi i N_i} \prod_{\langle i j\rangle } \mu^x_{ij} = 1,
%\label{constraint}
%\eea
%where the product runs over all the nearest-neighbors of the site $i$. These constraints relate the divergence of the gauge electric field, $\mu^x_{ij}$, to the gauge charge at every site - i.e., they impose a $\mathbb{Z}_2$ version of Gauss's law.

This discussion therefore leads to the schematic phase diagram of $\mathcal{H}_1$ shown
in Fig.~\ref{fig:xy}. Further details on the structure of the new phase with $\mathbb{Z}_2$ topological order appear
in Sections~\ref{sec:xy} and~\ref{sec:visons}. The model~(\ref{H1}) can support additional phases not shown in 
Fig.~\ref{fig:xy}, which we comment on in 
Appendix~\ref{app:phases}.

One important distinction between $\mathcal{H}_1$ and previous studies of fractionalization in doped Mott insulators 
\cite{TSMPAF00,LeeWenRMP,FG04,SS09}
is worth noting here. In all previous works, the Hamiltonian is 
presented in terms of emergent, fractionalized spinon and chargon degrees
of freedom. The electron operator does not appear explicitly in the Hamiltonian, but is described as a composite operator. 
In contrast, in our model $\mathcal{H}_1$ we have fractionalized the order parameter only into the spinons, while
retaining the bare electron operator in the Hamiltonian. 
In our approach, it is the chargon, rather than the electron, which appears as a composite particle,
as a bound state of the electron and the spinon in (\ref{psidef}).
We believe this difference in perspective is important, and that it leads to an efficient and controlled description of the
metallic states observed in the cuprates. 

We also comment here on an important subtlety in the structure of the metallic phase with $\mathbb{Z}_2$ topological order.
We have given arguments above on the appearance of reconstructed pocket Fermi surfaces of the $\psi_\pm$ chargons
in this phase. However, as we will see in our computations below, there remains a strong residual attraction 
\cite{PAL89,RKK07,RKK08,Punk15} between the chargons
and the spinons due to the hopping terms, $t_{ij}$ in $H_c$. Because of this attraction, it is possible that some or all of the chargons
form bound states with the spinons, leading to a pocket Fermi surface of electron-like quasiparticles with charge $e$ and 
spin $S_z = \pm 1/2$. If all of the chargons undergo this bound-state formation, then we obtain a FL* 
metal \cite{TSSSMV03,YQSS10,Mei12,Punk15,AT16,GCC16,LSW16}. 
Microscopic details
of the Hamiltonian will determined whether we obtain an ACL, or FL*, or an intermediate phase with co-existing chargon and electron 
Fermi surfaces \cite{RKK08}---the charge transport properties of all these phases are expected to be very similar, and so we will not
focus much on the distinction here. Note that, in the discussion above, 
even if all chargons bind with spinons to form electrons, we do not obtain
back the large Fermi surface Fermi liquid, but obtain FL*: this is because the Fermi surface size of the chargons is $p$ and not $1+p$.
Thus, an important feature of our analysis is that 
the operations of binding fermions to spinons, and of Fermi surface reconstruction, do {\em not\/} commute.

We note that a model closely related to $\mathcal{H}_1$ was studied by Grover and Senthil \cite{TGTS10} in the context of a two-band
Kondo-Heisenberg model. However, their interest was limited to the regime accessible perturbatively in an expansion in $\lambda$ (in our 
notation),
and they did not obtain the analog of the 
topological phase C in Fig.~\ref{fig:xy} with reconstructed Fermi surfaces. 
In doped Mott insulators, the coupling $\lambda \sim U$, the on-site Mott-Hubbard repulsion, and so $\lambda$ is the largest
energy scale in the Hamiltonian. We will work throughout in the large $\lambda$ limit, and will see below that the topological
phase appears in a regime inaccessible in a small $\lambda$ expansion.

The analysis of this paper will neglect superconductivity. But it should be noted that all the metallic 
phases of Fig.~\ref{fig:xy} are expected to superconduct at low $T$ \cite{SGTB15,LWYL15}.

The outline of the remainder of the paper is as follows.
Section~\ref{sec:xy} will analyze the properties of phase C of the easy-plane model $\mathcal{H}_1$ in a strong
coupling expansion. Section~\ref{sec:visons} will discuss the topological order and dynamics of visons in
$\mathcal{H}_1$. We will generalize the results to various cases in models with full SU(2) spin rotation invariance
in Section~\ref{sec:su2}. Section~\ref{sec:conc} will summarize our results and discuss the nature of the phase transitions
in Fig.~\ref{fig:xy}.

\section{Strong coupling expansion with  topological order}
\label{sec:xy}

This section will present a strong coupling analysis of the topological state C of the easy-plane
model $\mathcal{H}_1$ in Fig.~\ref{fig:xy}. The conventional metals A and B have the same properties 
as those of the phases of the SDW theory $H_{\rm sdw}$, and so don't need further discussion here.
Our analysis will establish the stability of a metallic phase with topological order, and also determine its excitation
spectrum in a limiting regime.

As the conventional phases appear in the limit of small $K$, we will study here the complementary $K \rightarrow \infty$ regime.
At $K = \infty$, we can work in the gauge $\mu^z_{ij} = 1$ everywhere.  So the model of interest in Eq.~(\ref{H1}) reduces to
the gauge-fixed Hamiltonian
\bea
\mathcal{H}_1^\prime &=& - \sum_{i,j} \left(t_{ij} + \mu \delta_{ij} \right) c_{i \alpha}^\dagger c_{j \alpha}
 - \lambda \sum_i \eta_i \left[ e^{- i \theta_i} c_{i \uparrow}^\dagger c_{i \downarrow} 
+ e^{ i \theta_i} c_{i \downarrow}^\dagger c_{i \uparrow} \right] \nn
&~&~~~~-  \sum_{i< j} J_{ij} \cos \left((\theta_i - \theta_j )/2\right) + 4\Delta \sum_i N_i^2 ,
\label{H1p}
\eea
describing fermions coupled to a XY rotor model. Note the crucial and only difference from the conventional
SDW theory $H_{\rm sdw}$ in Eq.~(\ref{HY0}): the $J_{ij}$ terms now involve couplings between the spinon field
$e^{i \theta/2}$, rather than the XY order parameter $e^{i \theta}$. Consequently, the rotor states on each
site have $N_i$ quantized in steps of 1/2, and the spin $S_z$ can be half-integer or integer.

This section will describe a strong-coupling analysis of $\mathcal{H}_1^\prime$ in which the on-site
terms are much larger than the off-site terms {\em i.e.\/} 
\beq
\lambda, \Delta \gg |t_{ij}|, |J_{ij}| . \label{sclimit}
\eeq
We will show that in this limit at $p=0$, $\mathcal{H}_1^\prime$ realizes a Mott insulator with a spin liquid
ground state. Furthermore, the spin liquid has odd $\mathbb{Z}_2$ topological order with gapped bosonic spinon and 
fermionic chargon
excitations: `odd' refers to the presence of unit $\mathbb{Z}_2$ background charge on each site of the Mott 
insulator \cite{RJSS91,MVSS99,TSMPAF00,MSF02};
the vison excitations have infinite energy at $K=\infty$, and will be considered further in Section~\ref{sec:visons}.
Moving towards $p>0$, we occupy the lowest energy fermonic holon states and obtain a ACL metal with odd
$\mathbb{Z}_2$ topological order. 

We will also describe the spectrum of states with one chargon and one spinon
excitation above the Mott insulator. Note that these states have total charge $e$ and spin $S_z = 1/2$, and so have
the same quantum numbers as the electron. We will find that along with the scattering states in which the holon
and spinon are well separated from each other, 
there is an electron-like bound state below the scattering continuum whose dispersion
we shall compute.
Because of the large spinon gap in the present strong coupling limit, 
this holon-spinon bound state is well above the band of fermionic holon states. 
However, away from the strong-coupling limit it is clearly possible that the 
bound state becomes the lowest energy charged fermionic state  \cite{RKK07,RKK08}. 
Then, at $p>0$, these states will
be occupied, leading to Fermi surfaces with electron-like quasiparticles. Such Fermi surfaces can co-exist with a holon
Fermi surfaces (leading to the `holon-hole' metal of Ref.~\onlinecite{RKK08}), or they can exist by themselves
in a $\mathbb{Z}_2$-FL* state.

\subsection{Single-site eigenstates}
We begin the  study of $\mathcal{H}_1^{\prime} $ in the limit (\ref{sclimit}) by defining the on-site Hamiltonian
\beq
H_0 = \sum_i \left(-\mu c_{i \alpha}^\dagger c_{i \alpha} - \lambda \, \eta_i \left[ e^{- i \theta_i} c_{i \uparrow}^\dagger c_{i \downarrow}
+ e^{ i \theta_i} c_{i \downarrow}^\dagger c_{i \uparrow} \right] + 4\Delta \sum_i N_i^2 \right)
\eeq

It is easy to determine all the eigenstates of $H_0$. In this subsection, we drop the site index, $i$. 
We denote the state with $N=0$ (no spinons) and no electrons as $\left| 0 \right\rangle$.
Then the eigenstate with $n$ spinons is 
\beq
\hat{N} e^{i n \theta/2} \left| 0 \right\rangle = \frac{n}{2} e^{i n \theta/2} \left| 0 \right\rangle.
\eeq
The empty electron state is implicit in $\left| 0 \right\rangle$, and all electrons will be indicated
below by creation operators acting on $\left| 0 \right\rangle$.
Notice that the $H_0$ conserves $n$ modulo 2, and states with $n$ odd carry the $\mathbb{Z}_2$ gauge charge.
The state $e^{i n \theta/2} \left| 0 \right\rangle$ carries spin $S_z = n/2$.
Important low-lying states are:\\
({\em i\/}) \underline{Mott insulator}\\ 
This is the state
\beq
\left| G \right\rangle = \frac{1}{\sqrt{2}} \left[\left(c_\uparrow^\dagger e^{-i \theta/2} +\eta \, c_\downarrow^\dagger e^{i \theta/2} \right) \left| 0 
\right\rangle \right] \quad , \quad E_G = - \lambda + \Delta - \mu. \label{mott1}
\eeq
This state carries no total spin, $S_z = 0$. It also carries $\mathbb{Z}_2$ gauge charge, and so the $\mathbb{Z}_2$ gauge theory is `odd' 
\cite{RJSS91,MVSS99,TSMPAF00,MSF02}, provided $E_G$ is the lowest energy state at half filling.\\
({\em ii\/}) \underline{Spinons}\\ 
These are doubly-degenerate states with $S_z = \pm 1/2$. The $S_z = +1/2$ state is
\beq
\left| \uparrow \right\rangle = \left( a \, c_{\uparrow}^{\dagger} + \eta b \, c_{\downarrow}^\dagger e^{2 i \theta/2} \right) \left|0 \right\rangle 
\quad; \quad E_s = -\mu + 2 \Delta - \sqrt{\lambda^2 + 4\Delta^2},
\eeq
where $(a,b)$ is an eigenvector of the matrix
\beq
\left( \begin{array}{cc} 0 & -\lambda  \\ -\lambda  & 4 \Delta \end{array} \right)
\eeq
with eigenvalue $2 \Delta - \sqrt{\lambda^2 + 4 \Delta^2}$, and similarly 
\beq
\left| \downarrow \right\rangle = \left( a \, c_{\downarrow}^{\dagger} + \eta b \, c_{\uparrow}^\dagger e^{-2 i \theta/2} \right) \left|0 \right\rangle 
\quad; \quad E_s = -\mu + 2 \Delta - \sqrt{\lambda^2 + 4\Delta^2}.
\eeq
Relative to the Mott insulator, the spinon carries no electromagnetic charge,
and a $\mathbb{Z}_2$ gauge charge. We want to be in a regime where the Mott insulator has a lower 
energy than the spinon, and so we require
\beq
E_G < E_s \quad \Rightarrow \quad \lambda > 3 \Delta /2.
\eeq
The spin gap, $\Delta_s$, of the Mott insulator is
\beq
\Delta_s = E_s - E_G = \Delta + \lambda - \sqrt{\lambda^2 + 4 \Delta^2}.
\eeq
({\em iii\/}) \underline{Holon}\\ 
This is simply the empty state $|0 \rangle$, with energy $E_{hn} = 0$. Relative to the Mott insulator, this state has
$S_z = 0$, $+e$ electromagnetic charge, and a non-zero $\mathbb{Z}_2$ gauge charge.\\
({\em iv\/}) \underline{Doublon}\\ 
This is the  state $c_\uparrow^\dagger c_\downarrow^\dagger |0 \rangle$, with energy $E_{dn} = -2 \mu$. Relative to the Mott insulator, this 
state has
$S_z = 0$, $-e$ electromagnetic charge, and a non-zero $\mathbb{Z}_2$ gauge charge.\\
({\em v\/}) \underline{Holes}\\
These are the doubly degenerate states $e^{\pm i \theta/2} \left| 0 \right\rangle$ with energy $E_h = \Delta$. Relative to the Mott insulator, 
they carry electromagnetic charge $+e$ and zero $\mathbb{Z}_2$ gauge charge. They have spin $S_z = \pm 1/2$. We 
can now examine the energy difference between a pair of sites with hole+Mott insulator and a pair with holon+spinon
\beq
E_h + E_G - E_{hn} - E_s = 
- \lambda + \sqrt{\lambda^2 + 4 \Delta^2} > 0. \label{eh}
\eeq
So a hole is unstable to decay into a holon and a spinon in the strong-coupling expansion of Eq.~(\ref{sclimit}). 
Note that, at $\lambda \gg \Delta$, this 
energy difference can become small.\\
({\em vi\/}) \underline{Electrons}\\
These are the doubly degenerate states $c_\uparrow^\dagger c_{\downarrow}^\dagger e^{ \pm i \theta/2} \left| 0 \right\rangle$ with energy 
$E_e = -2 \mu + \Delta$. Relative to the Mott insulator, they carry electromagnetic charge $-e$ and zero $\mathbb{Z}_2$ gauge charge. 
They have spin $S_z = \pm 1/2$. 
The condition for the instability of an electron state is 
\beq
E_e + E_G - E_{dn} - E_s = 
- \lambda + \sqrt{\lambda^2 + 4 \Delta^2} > 0,
\eeq
which is the same as (\ref{eh}).

For subsequent analysis, it is useful to introduce the canonical fermion operators of the holon (this is a linear combination
of the operators in (\ref{psidef}))
\beq
\psi = \frac{1}{\sqrt{2}} \left( e^{i \theta/2} c_{\uparrow} + \eta \, e^{-i \theta/2} c_{\downarrow} \right) \label{psidef2}
\eeq
Note that $\psi$ is the holon creation operator, {\em i.e\/} a holon is an empty state in a filled $\psi$ band,
\beq
\left| G \right\rangle = \prod_i \psi_i^\dagger \left| 0 \right\rangle, \label{Mott}
\eeq
which is the Mott insulator.

We also introduce the fermions
\bea
\Phi_\uparrow^\dagger &=& a \, c_{\uparrow}^{\dagger} + \eta b \, c_{\downarrow}^\dagger e^{ i \theta} \nn
\Phi_\downarrow^\dagger &=& a \, c_{\downarrow}^{\dagger} + \eta b \, c_{\uparrow}^\dagger e^{- i \theta}
\eea
Then the spinon creation operator is the boson
\beq
b_{\alpha}^\dagger  = \Phi_{\alpha}^\dagger \psi
\label{spinonOp}
\eeq
This creates the spinon excitation via $b_\alpha^\dagger \left| G \right\rangle$.

\subsection{Effective holon Hamiltonian}

Now we move beyond the single-site Hamiltonian, and examine the influence of the multisite terms on the
single holon excitation above the Mott insulator.

First, we note the on-site  Hamiltonian
\beq
H_{h0} = \sum_{i}  \left(- \mu + \Delta - \lambda \right) \psi_{i}^\dagger \psi_{i}.
\eeq
which describes the on-site energy of the holon states.

Next, we include the hopping terms $t_{ij}$ and $J_{ij}$. We perform a canonical transformation to eliminate the $\theta/2$ excitations to 
obtain an effective Hamiltonian for the holons.
This transformation should be performed around single particle excitations of the band insulator of $\psi$, which is the Mott insulator 
$\left| G \right\rangle$.
For this transformation, it is convenient go back to the original $c$ fermion formulation. 
We make a list of all states among a pair of sites, $1,2$,
which are important to second order perturbation theory in $t$,$J$ with a total charge of $e$ and a total $S_z$ of 0; there turn
out to be 12 such states
\bea
c_{1 \uparrow}^\dagger e^{-i \theta_1/2} \left| 0 \right\rangle \quad , \quad
 c_{1 \downarrow}^\dagger e^{i \theta_1/2} \left| 0 \right\rangle \quad &,& \quad
 c_{2 \uparrow}^\dagger e^{-i \theta_2/2} \left| 0 \right\rangle \quad , \quad
 c_{2 \downarrow}^\dagger e^{i \theta_2/2} \left| 0 \right\rangle \nn
 c_{2 \uparrow}^\dagger e^{-i \theta_1/2} \left| 0 \right\rangle \quad , \quad
 c_{2 \downarrow}^\dagger e^{i \theta_2 -i \theta_1/2} \left| 0 \right\rangle \quad &,& \quad
 c_{2 \downarrow}^\dagger e^{i \theta_1/2} \left| 0 \right\rangle \quad , \quad
 c_{2 \uparrow}^\dagger e^{-i \theta_2 +i \theta_1/2} \left| 0 \right\rangle \nn 
 c_{1 \uparrow}^\dagger e^{-i \theta_2/2} \left| 0 \right\rangle \quad , \quad
 c_{1 \downarrow}^\dagger e^{i \theta_1 -i \theta_2/2} \left| 0 \right\rangle \quad &,& \quad
 c_{1 \downarrow}^\dagger e^{i \theta_2/2} \left| 0 \right\rangle \quad , \quad
c_{1 \uparrow}^\dagger e^{-i \theta_1 +i \theta_2/2} \left| 0 \right\rangle 
\eea
Each site in all of these states is limited to have a spin of $S_z = 0, \pm 1/2$.
A conventional computation then eliminates the last 8 of these states to yield the effective holon Hamiltonian.

To leading order in $J/\lambda$, only hopping within the same sublattice contributes, and the effective holon Hamiltonian turns out to be
\beq
H_h = H_{h0} + H_{h1} \label{Hh}
\eeq
with
\beq
H_{h1} =  - \sum_{i<j,\, n.n.n.} t_2  J_2 Z \left(\psi_{i}^\dagger \psi_{j} + \psi_{j}^\dagger \psi_{i} \right)  - \sum_{i<j,\, n.n.n.n.} t_3 J_3 Z 
\left(\psi_{i}^\dagger \psi_{j} + \psi_{j}^\dagger \psi_{i} \right). 
\eeq
where
\beq
Z= \frac{(\lambda+2\Delta)}{2 \Delta \lambda}. \label{Zh}
\eeq
Here, $n.n.n.$ and $n.n.n.n.$ stand for second-and third-neighbor sites, respectively, and the renormalization factor $Z$ is related
to the $Z_{ij}$ mentioned in Section~\ref{sec:fracorder}.
One feature of $H_{h1}$ is that the holons on the two sublattices don't mix with each other; {\em i.e.\/}, hopping between same-sublattice 
sites on the square lattice is forbidden.
This is an exact property of this model
due to a symmetry of the XY model: the holon operator in (\ref{psidef2}) is odd or even under the spin inversion $S_z \rightarrow - S_z$, $
\theta \rightarrow - \theta$ on the two sublattices. 

\subsection{Effective spinon Hamiltonian}
\label{sec:spinons}

Next we examine the hopping of the single spinon excitations above the Mott insulator. In this case, the computation is simpler
than the holon case, and the
the spinon Hamiltonian is easily obtained
by first-order perturbation theory:
\beq
H_s =  \sum_i \left(\Delta + \lambda - \sqrt{\lambda^2 + 4 \Delta^2} \right) b_{i \alpha}^\dagger b_{\alpha} - \frac{(a+b)^2}{2} \sum_{i<j} J_{ij} 
\left( b_{i \alpha}^\dagger b_{j \alpha} + b_{j \alpha}^\dagger b_{i \alpha} \right).
\eeq

\subsection{Holon and spinon bound state}

We are now ready to consider the states with both a holon and a spinon present. 
The most important coupling between them appears already 
at first order in $t_{ij}$ when the holon and the spinon exchange positions. The matrix element for this is easily computed and leads
to the Hamiltonian

\bea
H_{hs1} &=&  - \sum_{i<j} a^2 t_{ij}  \left( \Phi_{i \alpha}^\dagger \Phi_{j \alpha} +  \Phi_{j \alpha}^\dagger \Phi_{i \alpha} \right) \nn 
\label{Hcs2}
&=&  - \sum_{i<j} a^2 t_{ij}  \left( \psi_i^\dagger \psi_j b_{i \alpha}^\dagger b_{j \alpha} +  \psi_j^\dagger \psi_i b_{j \alpha}^\dagger b_{i 
\alpha} \right).
\eea

Introducing the holon operators $h = \psi^\dagger$, and collecting all terms, the Hamiltonian acting on the Hilbert space of one holon and 
one spinon is
\bea
H_{hs} &=& \sum_{i} \left[ \left( \mu - \Delta + \lambda \right) h_{i}^\dagger h_{i} + \left(\Delta + \lambda - \sqrt{\lambda^2 + 4 \Delta^2} \right) 
b_{i \alpha}^\dagger b_{\alpha} \right] \nn
&~& + \sum_{i<j,\, n.n.n.} \frac{t_2  J_2 (\lambda + 2 \Delta)}{2 \Delta \lambda} \left(h_{i}^\dagger h_{j} + h_{j}^\dagger h_{i} \right)  + 
\sum_{i<j,\, n.n.n.n.}\frac{t_3  J_3 (\lambda + 2 \Delta)}{2 \Delta \lambda} \left(h_{i}^\dagger h_{j} + h_{j}^\dagger h_{i} \right) \nn
&~& - \frac{(a+b)^2}{2} \sum_{i<j} J_{ij} \left( b_{i \alpha}^\dagger b_{j \alpha} + b_{j \alpha}^\dagger b_{i \alpha} \right)
+ \sum_{i<j} a^2 t_{ij}  \left( h_j^\dagger h_i b_{i \alpha}^\dagger b_{j \alpha} +  h_i^\dagger h_j b_{j \alpha}^\dagger b_{i \alpha} \right) 
\label{Hhs}
\eea
along with a hard-core constraint which prevents the holon and spinon from residing on the same site. 

A number of interesting features of $H_{hs}$ deserve notice. In the strong-coupling limit of (\ref{sclimit}), the holon+spinon
states are at larger energy than the energy of a single holon. However, there is an attractive interaction between the holon
and spinon ($\sim t$) which is parametrically larger than the bandwidth of the holon ($\sim t J /\lambda$). This implies that there
will be a clear separation between the energy of the holon+spinon bound state and the bottom of the holon-spinon continuum,
and this will be evident from our numerical results below.
Although such an electron-like bound state does form in the strong coupling limit, its energy remains higher than
the energy of the single holon band. This implies that doping the Mott insulator will lead to a Fermi surface of holons only, 
realizing a $\mathbb{Z}_2$-ACL metal in the limit (\ref{sclimit}). However, as we move away from (\ref{sclimit}), the conditions
are favorable for the holon+spinon bound state to become the lowest energy charged fermion, and doping will then
lead to a $\mathbb{Z}_2$-FL* metal. In particular, FL* is favored in the limit $\lambda \gg \Delta \gg |t_{ij}| \gg |J_{ij}|$,
when the on-site energy cost of the hole state in Eq.~(\ref{eh}), which is $\sim \Delta^2/\lambda$, can be compensated by an energy gain $\sim - |t_{ij}|$ from the kinetic energy of the hole.
\begin{figure}
\begin{center}
\includegraphics[height=8.6cm]{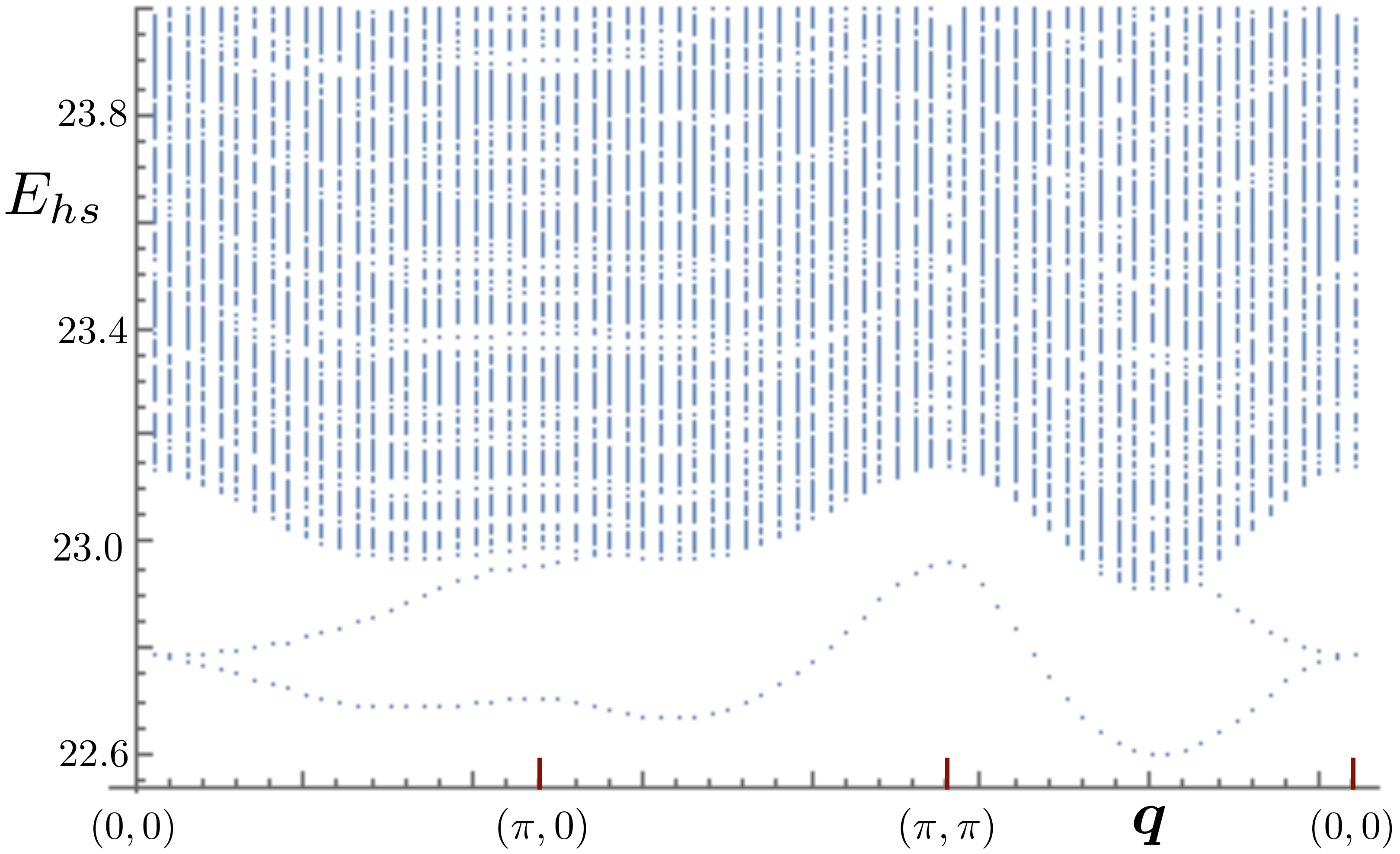}
\end{center}
\caption{Lowest energy eigenvalues of $H_{hs}$ with one holon and one spinon with total momentum ${\bf q}$ on a lattice of size
$48 \times 48$. There are $48^2=2304$ eigenvalues at each ${\bm q}$, and eigenvalues above $E_{hs} = 24.0$ are not shown. Note the 
bound states (which have charge $e$ and
spin $S_z = \pm 1/2$) below the two particle continuum. The parameter values are $\lambda = 30.0$, $\Delta = 8.0$, $t_1=3.0$, $t_2 = 
2.0$, $t_3 = 2.0$,
$J_1 = 0.6$, $J_2 = 0.1$, $J_3 = 0.1$, and $\mu=0$. The energy levels shift uniformly with changes in $\mu$,
and the bound state will form a pocket with electron-like quasiparticles for large enough $\mu$. There is also \cite{RKK08} a Fermi surface
of chargons associated with the single holon states, which are not shown above.
}
\label{fig:boundstate}
\end{figure}
\begin{figure}
\begin{center}
\includegraphics[height=8cm]{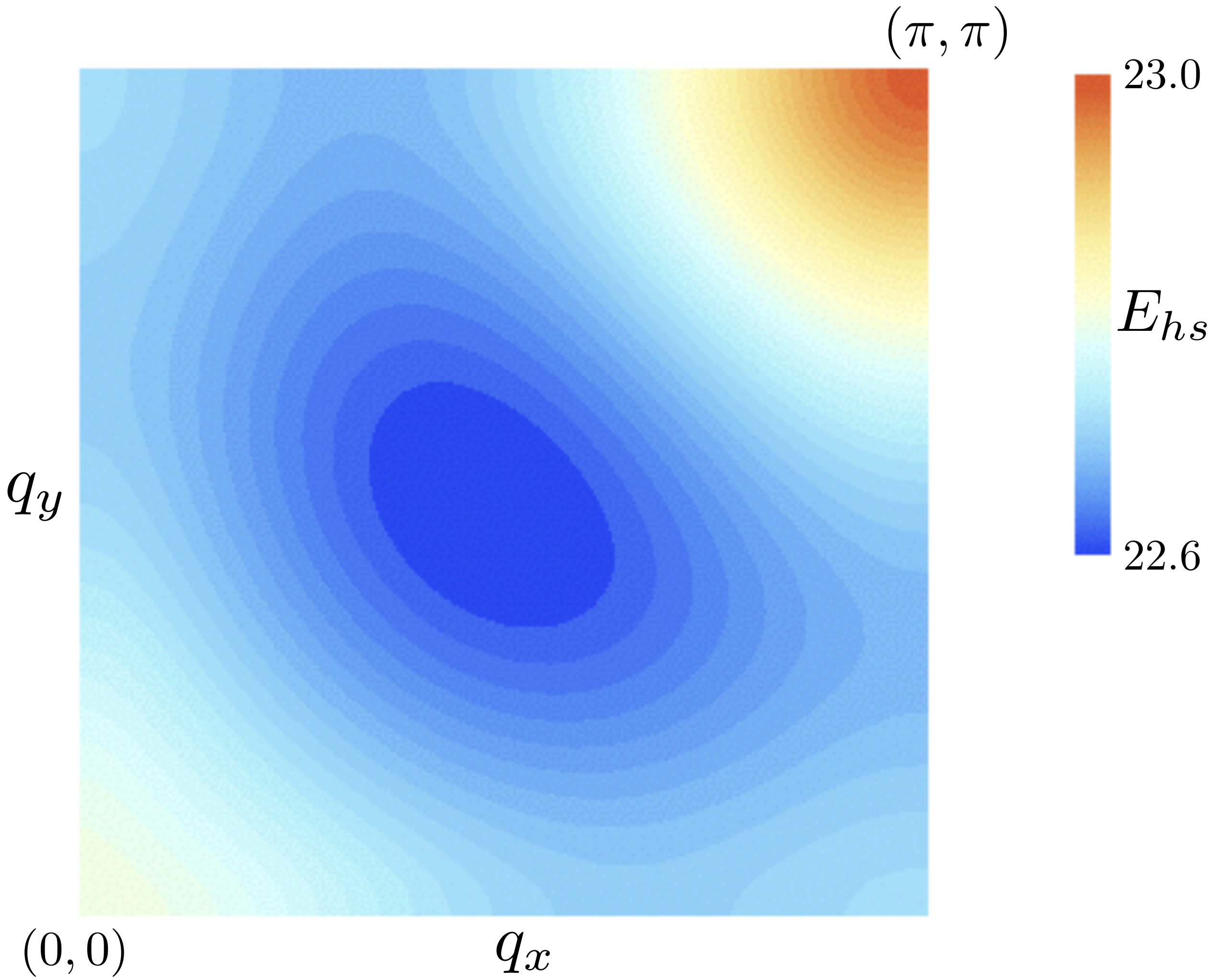}
\end{center}
\caption{Colour density plot of the energy of the lowest holon-spinon bound state in Fig.~\ref{fig:boundstate}. Parameter values are the same. Note that there is no particular symmetry of the bound state dispersion associated with antiferromagnetic Brillouin
zone: the minimum of the dispersion is not exactly at $(\pi/2, \pi/2)$. In contrast, the holon dispersion given by 
Eq.~(\ref{Hh}) does have a minimum at $(\pi/2, \pi/2)$.}
\label{fig:holepocket}
\end{figure}

We now establish the above assertions by an exact diagonalization study of $H_{hs}$ in (\ref{Hhs}) in the sector with one holon
and one spinon. This is most conveniently carried out in the momentum space Hamiltonian
\bea
H_{hs} &=& \sum_{\bk} E_h (\bk) h_{\bk}^\dagger h_{\bk} +  \sum_{\bk} E_b (\bk) b_{\bk\alpha}^\dagger b_{\bk\alpha}
+ \frac{1}{L^2} \sum_{\bk,\bk', \bq} V (\bk+\bp+\bq) h_{\bk + \bq}^\dagger b_{-\bk,\alpha}^\dagger b_{-\bp,\alpha} h_{\bp + \bq} \nn
E_h (\bk) &=&  \left( \mu - \Delta + \lambda \right) +  \frac{t_2  J_2 (\lambda + 2 \Delta)}{2 \Delta \lambda} 
\left( 4 \cos (k_x) \cos (k_y) \right) + \frac{t_3  J_3 (\lambda + 2 \Delta)}{2 \Delta \lambda} \left( 2 \cos (2k_x) + 2 \cos (2 k_y) \right) \nn
E_b (\bk) &=& \left(\Delta + \lambda - \sqrt{\lambda^2 + 4 \Delta^2} \right) - \frac{(a+b)^2}{2} \left[ 2 J_1 (\cos (k_x) + \cos (k_y)) +  4 J_2 
\cos (k_x) \cos (k_y)) \right. \nn
&~&~~~~~~~~ \left. + 2 J_3 (\cos (2 k_x) + \cos (2k_y)) \right] \nn
V (\bk) &=& a^2 W + a^2 \left[ 2 t_1 (\cos (k_x) + \cos (k_y)) +  4 t_2 \cos (k_x) \cos (k_y)) + 2 t_3 (\cos (2k_x) + \cos (2k_y))\right] , 
\label{Hhs2}
\eea
where $W \rightarrow \infty$ is a large repulsive energy inserted to prevent the holon and spinon from occupying the same site.
We diagonalized Eq.~(\ref{Hhs2}) on a $L \times L$ lattice: after accounting for total momentum conservation, the matrix in the
holon+spinon subspace is of size $L^2 \times L^2$. Results for a convenient choice of parameters are shown in 
Figs.~\ref{fig:boundstate} and~\ref{fig:holepocket}.

\section{Dynamics of visons}
\label{sec:visons}

We now discuss aspects of the topological order of the Mott insulator described so far; the metallic phase C in Fig.~\ref{fig:xy} inherits the 
same topological order. These issues require us to return to the full gauge-invariant 
Hamiltonian in Eq.~(\ref{H1}), and to no longer
work with the large $K$ gauge-fixed version in Eq.~(\ref{H1p}).

Here, a very useful fact is that the Mott insulator in Eq.~(\ref{mott1}) and the $\mathbb{Z}_2$ gauge
theory of its bosonic spinon excitations are essentially identical to the theory of bosonic chargons presented by
Paramekanti and Vishwanath in Section~V.A of Ref.~\onlinecite{APAV04};
and the change from the spinon to chargon character of the bosons
makes essentially no difference to the topological analysis. 
The main observation is that the Mott insulator in Eq.~(\ref{mott1}) only has states in which the number $2N_i$ equals
$\pm 1$ on every site. (We can view the deviation of $2N_i$ from $\pm 1$ as a measure of the number of spinon excitations.
Or by a slight abuse of language, we identify $2N_i$ as the number of spinons `in the ground state'.) This observation 
motivates a return to the Hamiltonian in Eq.~(\ref{H1})
at $p=0$, from which we  
integrate out the gapped $c_{i \alpha}$ electronic excitations at large $\lambda$, when there is a large energy
cost to deviations of the number, $2N_i$, from $\pm 1$. So, we obtain an effective Hamiltonian of the form
\bea
\widetilde{H}_{\theta,\mathbb{Z}_2} &=& -  \sum_{i< j} J_{ij} \mu^{z}_{ij} \cos \left((\theta_i - \theta_j )/2\right)  
- g \sum_{\langle ij \rangle} \mu^x_{ij} - K  \sum_{\square} \left[\prod_{\square} \mu^z_{ij}\right] \nn
&~&~~~~~~~~+ \widetilde{\Delta} \sum_i (4N_i^2 - 1)^2,
\label{HZ2}
\eea
where the $\widetilde{\Delta}$ term is a phenomenological representation of the energy cost for deviation of 
the `spinon' number $2N_i$ from 
$\pm 1$. Eq.~(\ref{HZ2}) is essentially the Hamiltonian  $H_A (\mathcal{I}^\ast)$ of Ref.~\onlinecite{APAV04}. 
All of their arguments
associated with momentum balance in the presence of flux insertion in a torus geometry go through unchanged:
so we have established the existence of the needed Mott insulator described by an odd $\mathbb{Z}_2$ gauge
theory \cite{RJSS91,MVSS99,TSMPAF00,MSF02}, which can then act as a parent for metallic states with Fermi surfaces of size $p$ 
\cite{SSDC16}.

As argued in Ref.~\onlinecite{APAV04}, we can proceed a step further and also integrate out the gapped
spinon excitations from Eq.~(\ref{HZ2}). Then, we obtain a pure $\mathbb{Z}_2$ gauge theory
\beq
\widetilde{H}_{\mathbb{Z}_2} =  
- g \sum_{\langle ij \rangle} \mu^x_{ij} - K  \sum_{\square} \left[\prod_{\square} \mu^z_{ij}\right], \label{HZ2v}
\eeq
where the `spinons' in the ground state can be accounted for by an  `odd' constraint on every site $i$
derived from Eq.~(\ref{Gi})
\beq
\hat{\overline{G}}_i \equiv \prod_{j \in {\rm n.n.}(i)} \mu^x_{ij} = -1. \label{constv}
\eeq
This is the most convenient form of the theory to investigate the dynamics
of visons. Note that if we have obtained an effective theory of visons simply by integrating out the $c_{i \alpha}$ in a 
weak-coupling perturbation theory in $\lambda$, then we would have obtained the effective theory in Eq.~(\ref{HZ2v}), but with 
the opposite sign even constraint in Eq.~(\ref{constv}): this was the procedure used in Ref.~\onlinecite{TGTS10}. So the
important constraint in Eq.~(\ref{constv}) relies crucially on our focus on the large $\lambda$ physics, and the fact that at large $\lambda$
each electron binds a spinon into the state $\left| G \right\rangle$ in Eq.~(\ref{Mott}).

\begin{figure}
\begin{center}
\includegraphics[height=6cm]{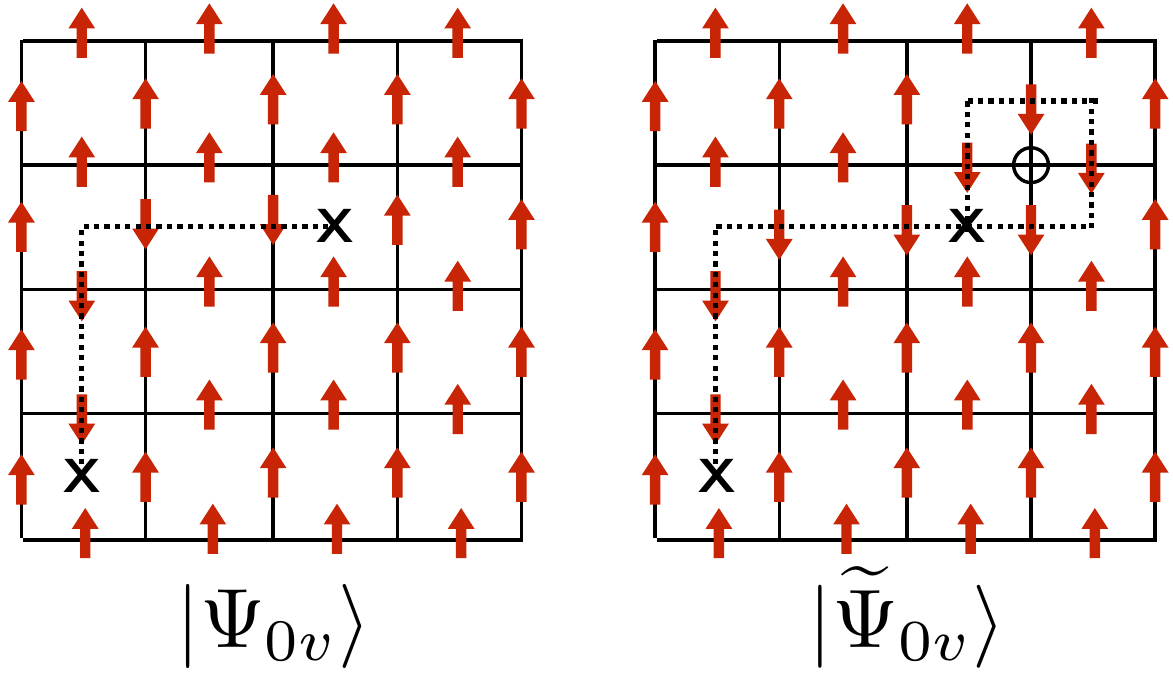}
\end{center}
\caption{A gauge-fixed state, $|\Psi_{0v} \rangle$ of two visons marked with the X's. The dotted line connects links
with $\mu^z_{ij} = -1$. The state $|\widetilde{\Psi}_{0v} \rangle$ is obtained after the top vison encircles the 
site, $j$, marked with a circle.
}
\label{fig:vison}
\end{figure}

In the large $K$ limit, the ground state of $\widetilde{H}_{\mathbb{Z}_2}$ with the constraint
in Eq.~(\ref{constv}) is the same as an `odd' toric code model \cite{Kitaev03}.
Starting with the gauge-fixed ground state, $|\Psi_0\rangle$, used in Section~\ref{sec:xy} with all $\mu^z_{ij} = 1$,
we can obtain a ground state, $|\Psi_G \rangle$ obeying Eq.~(\ref{constv}) by summing over all combinations of gauge transformations applied to this state (normalized on a torus with an even number of sites)
\beq
|\Psi_G \rangle = \frac{1}{\sqrt{2}} \prod_i \frac{1}{\sqrt{2}} \left( 1 - \hat{\overline{G}}_i \right) |\Psi_0 \rangle .
\eeq
All the terms on the right hand side have $\prod_\square \mu^z_{ij}=1$ on every plaquette, and so minimize the energy at 
large $K$. Similarly, we can also obtain the vison excited states in the large $K$ limit from a gauge-fixed vison wavefunction. 
We show in Fig.~\ref{fig:vison} a two vison state, $|\Psi_{0v} \rangle$, with two plaquettes (each marked by
an X) which have $\prod_\square \mu^z_{ij}=-1$. 
The two vison state obeying Eq.~(\ref{constv}) is obtained by
\beq
|\Psi_{Gv} \rangle = \frac{1}{\sqrt{2}} \prod_i \frac{1}{\sqrt{2}} \left( 1 - \hat{\overline{G}}_i \right) |\Psi_{0v} \rangle .
\eeq
Unlike the toric code, these visons are not localized, and acquire a non-zero dispersion at linear order in a perturbation
theory in $g/K$. A crucial property of these mobile visons in a odd $\mathbb{Z}_2$ gauge theory is that
they experience a Berry phase of $\pi$ upon encircling any site of the square lattice \cite{RJSS91,MVSS99,TSMPAF00}. 
This is apparent from the state
$|\widetilde{\Psi}_{0v} \rangle$ in Fig.~\ref{fig:vison} in which the top vison has encircled the site, $j$, marked with 
the circle. Then, it is easy to show that vison states obeying Eq.~(\ref{constv}) satisfy
\beq
|\widetilde{\Psi}_{Gv} \rangle = \frac{1}{\sqrt{2}} \prod_i \frac{1}{\sqrt{2}} \left( 1 - \hat{\overline{G}}_i \right) |\widetilde{\Psi}_{0v} \rangle 
= \hat{\overline{G}}_j |\Psi_{Gv} \rangle = - |\Psi_{Gv} \rangle .
\eeq
This Berry phase leads to a double degeneracy in the vison spectrum at all momenta.
Alternatively, we can obtain an
effective Hamiltonian for the visons by applying a duality transformation to Eq.~(\ref{HZ2v}): this can be done
using the operator methods described by Kogut \cite{KogutRMP}, by the perturbative Berry phase computation 
of Ref.~\onlinecite{YHMPSS11},
or by path integral methods \cite{RJSS91,MVSS99,TSMPAF00}.
The result by any of these methods is a transverse field Ising model on the dual
lattice, with the Ising order representing the vison field operator.
The odd constraint in Eq.~(\ref{constv}) leads to $\pi$ flux per plaquette in the vison hopping matrix elements.
Such a model for visons was extended to FL* metals in recent work \cite{PCAS16}.

\section{Generalizations to SU(2) global spin symmetry}
\label{sec:su2}

All of our analysis so far has been restricted to the simplest case with only a U(1) global spin rotation symmetry, so that the 
SDW order is described by a XY order parameter. Now we consider the generalization to the physically important
case of full SU(2) spin rotation symmetry. There are now significant differences determined by the specific configuration
of the SDW order. In general, we can characterize a state with long-range SDW order
by the expectation value
\beq 
\left\langle c_{i\alpha}^\dagger \sigma^\ell_{\alpha\beta} c_{i\beta} \right\rangle
\sim \Phi_{x\ell} \, e^{ i {\bm K}_x \cdot{\bm r}_i } + \mbox{c.c.} + \Phi_{y\ell} \, e^{ i {\bm K}_y \cdot{\bm r}_i } + \mbox{c.c.}.
\label{norder}
\eeq
where $\sigma^\ell$ are the Pauli matrices, 
${\bm K}_{x,y}$ are ordering wavevectors along the $x$ and $y$ directions, and $\Phi_{x \ell}, \Phi_{y \ell}$ are
6 complex numbers determining the nature of the spin density order.  The quantum fluctuations of the $\Phi$
are controlled by a Landau free energy of the form \cite{ZDS02}
\bea
V(\Phi) &=& s (\left|\Phi_{x \ell}\right|^2
   + \left|\Phi_{y \ell}\right|^2)
   +\frac{u_1}{2} ( \left|\Phi_{x \ell}\right|^4
   + \left|\Phi_{y \ell}\right|^4 )
 + \frac{u_2}{2} ( \left|\Phi_{x \ell}^2\right|^2    + \left|\Phi_{y \ell}^2\right|^2 ) \nn
 &~&~~
   + w_1 \left|\Phi_{x \ell}\right|^2\left|\Phi_{y \ell}\right|^2 + w_2 \left|\Phi_{x \ell} \Phi_{y \ell}\right|^2
   + w_3 \left|\Phi_{x \ell}^* \Phi_{y \ell} \right|^2 + \ldots
\eea
Depending upon the relative values of the Landau parameters $w_{1,2,3}$, the 6 complex numbers $\Phi_{x \ell}, \Phi_{y \ell}$
can realize physically distinct types of spin density wave order which we consider separately in the subsections below.
Each type of order leads to a different routes towards taking the `square root' of the order parameter
into fractionalized spinor variables.

\subsection{Spiral order}
\label{sec:spiral}

The case most similar to the easy-plane order is when then SDW order has a spiral form at a wavevector
not equal to $(\pi, \pi)$. 
For simplicity, we consider the case with circular spiral spin correlations only along the wavevector ${\bm K}_x$; it is not difficult to extend 
the action below to also include the ${\bm K}_y$ direction. 
The case corresponds to (after an overall normalization)
\beq
\Phi_{x \ell} = n_{1\ell} + i n_{2 \ell} \label{Pn1}
\eeq
where $n_{1,2 \ell}$ are a pair of real orthonormal vectors
\beq
\sum_{\ell} n_{1 \ell}^2 = n_{2 \ell}^2 = 1 \quad , \quad \sum_{\ell} n_{1 \ell} n_{2 \ell} = 0.  \label{Pn2}
\eeq
The order parameter defined by Eqs.~(\ref{Pn1}) and (\ref{Pn2}) is a doublet of orthonormal 3-vectors, and this
is equivalent to the SO(3) manifold. 

To obtain the intermediate metallic state with topological order, we need to fractionalize the above order.
For the easy-plane case, we accomplished this by working with the square-root of the order parameter, $e^{i \theta/2}$.
Here, we need to introduce a bosonic complex spinor $z_\alpha$, representing the spinon excitation, which we take to be of unit length
\beq
|z_\uparrow|^2 + |z_\downarrow|^2 = 1. \label{zconst}
\eeq
Then the parametrizations in Eqs.~(\ref{Pn1},\ref{Pn2}) can be satisfied by the representation \cite{SSNR91,SSRMP}
\beq
\Phi_{x \ell} = \varepsilon_{\alpha\gamma} z_\gamma \sigma^{\ell}_{\alpha\beta} z_\beta \label{Phiz}
\eeq
Note that Eq.~(\ref{Phiz}) is invariant under the $\mathbb{Z}_2$ gauge transformation $z_\alpha \rightarrow -z_\alpha$,
and so we will again obtain here a $\mathbb{Z}_2$ gauge theory, similar to the easy-plane case (which corresponds
to $z_{\alpha} = (e^{-i \theta/2}, 0)$).

We can now write down our theory for the interplay between spiral SDW order and $\mathbb{Z}_2$ topological
order, which is analogous to the easy-plane Hamiltonian $\mathcal{H}_1$ in Eq.~(\ref{H1}),
\bea
\mathcal{H}_2 &=& H_c + H_{z,\mathbb{Z}_2} + H_{Y} \nn
H_{Y} &=& - \lambda\sum_i \left[ \,  \varepsilon_{\alpha\gamma} z_{i\gamma} \sigma^{\ell}_{\alpha\beta} z_{i\beta} \, e^{ i {\bm K}_x 
\cdot{\bm r}_i } + \mbox{c.c.}  \right]
\, c_{i\alpha}^\dagger {\sigma}^\ell_{\alpha\beta} c_{i\beta} \nn
H_{z,\mathbb{Z}_2} &=& -  \sum_{i< j} J_{ij} \, \mu^{z}_{ij} \left( z_{i\alpha}^\ast z_{j \alpha} + \mbox{c.c.} \right)
+ \Delta \sum_i \vec{L}_i^2 
- g \sum_{\langle ij \rangle} \mu^x_{ij} - K  \sum_{\square} \left[\prod_{\square} \mu^z_{ij}\right],
\label{H2}
\eea
where $H_c$ was defined in Eq.~(\ref{Hc}), and $\vec{L}_i$ are angular momenta of the O(4) rotor defined by (\ref{zconst}), analogous to $N_i$ for the easy-plane case. The on-site eigenstates of $\mathcal{H}_2$, in the limit of $K \rightarrow \infty$ and strong coupling as defined by Eq.~(\ref{sclimit}) , are described in Appendix \ref{app:OnSiteSpiral}.

The analog of the relationship between the chargon and electron operators in Eq.~(\ref{psidef}) transforming to the rotating reference
frame now becomes the SU(2) rotation
\beq
\left( \begin{array}{c} \psi_+ \\ \psi_- \end{array} \right) =
\left( \begin{array}{cc}  
z_\uparrow^\ast & z_\downarrow^\ast \\
- z_\downarrow & z_\uparrow
\end{array} \right) 
\left( \begin{array}{c} c_\uparrow \\ c_\downarrow \end{array} \right).
\label{czpsi}
\eeq
In terms of the chargon operators, the Yukawa term, $H_{Y}$ in Eq.~(\ref{H2}) takes the simple form \cite{DCSS15b}
\beq
H_{Y} = - 2 \lambda \sum_i \left[ \,  \psi_{i+}^\dagger \psi_{i-} \, e^{ i {\bm K}_x \cdot{\bm r}_i } + \mbox{c.c.}  \right],
\label{HY2}
\eeq
which is the analog of Eq.~(\ref{Hsdw2}). So analogous to the easy-plane case,
the $\psi$ fermions move in a background of spatially uniform spiral order. 

The subsequent discussion is a close parallel to that described above for the easy-plane case. We expect
a phase diagram very similar to that in Fig.~\ref{fig:xy}, with the $e^{i \theta}$ order parameter replaced by
$\Phi_{x\ell}$. One difference is that the interpretation of the phase
transitions in terms of vortex proliferation now needs some modification, as the order parameter is no longer XY-like but
takes values in SO(3).
Such an order parameter does have $\mathbb{Z}_2$ vortices, associated
with the homotopy group 
$\pi_1 \left( {\rm SO}(3) \right) = \mathbb{Z}_2$. So the topological phase C is now associated with the suppression
of such $\mathbb{Z}_2$ vortices, which become gapped excitations identified with visons. In contrast, the Fermi liquid
phase A has proliferating $\mathbb{Z}_2$ vortices. 

\subsection{N\'eel order}
\label{sec:neel}

Next, we consider the case of two-sublattice collinear antiferromagnetism. For the cuprates, this corresponds in Eq.~(\ref{norder})
to the wavevectors ${\bm K}_x = {\bm K}_y = (\pi, \pi)$ and real $\Phi_{x\ell} = \Phi_{y\ell}$, and is applicable
to the electron-doped compounds.

In this case, the order parameters are related to a single, real vector $\Phi_{x\ell} = \Phi_{y\ell} = n_{\ell}/4$ obeying
\beq
\sum_{\ell} n_{\ell}^2 = 1. \label{n21}
\eeq
We fractionalize this vector by
\beq
n_{\ell} = z_\alpha^\ast \sigma^\ell_{\alpha\beta} z_\beta,
\eeq
which leaves a U(1) gauge invariance under $z_\alpha \rightarrow e^{i f} z_\alpha$.
So now, our generalization of the $\mathbb{Z}_2$ gauge theory $\mathcal{H}_2$ in Eq.~(\ref{H2})
is a U(1) gauge theory for SDW order in metals:
\bea
\mathcal{H}_3 &=& H_c + H_{z,U(1)} + H_{Y} \nn
H_{Y} &=& - \lambda\sum_i \eta_i \, z_{i\alpha}^\ast \sigma^{\ell}_{\alpha\beta} z_{i\beta} \, c_{i\alpha}^\dagger {\sigma}^\ell_{\alpha\beta} 
c_{i\beta} \nn
H_{z,U(1)} &=& -  \sum_{i< j} J_{ij} \, \left( z_{i\alpha}^\ast e^{i A_{ij}} z_{j \alpha} + \mbox{c.c.} \right)
+ \Delta \sum_i \vec{L}_i^2 
+ g \sum_{\langle ij \rangle} E_{ij}^2  - K \sum_{\square}\cos \left( \sum_{\square} A_{ij} \right) ,
\label{H3}
\eea
where $A_{ij}$ is a compact U(1) gauge field on the links of the square lattice, and $E_{ij}$ is the canonically conjugate electric
field. The on-site eigenstates of $\mathcal{H}_3$, in the limit of $K \rightarrow \infty$ and strong coupling as defined by Eq.~(\ref{sclimit}) , are described in Appendix \ref{app:OnSiteNeel}.

The analysis proceeds as in Section~\ref{sec:spiral}. We transform to the rotation reference frame in terms of chargons
as in Eq.~(\ref{czpsi}), and the Yukawa coupling for the chargons is \cite{SS09}
\beq
H_{Y} = - \lambda \sum_i \eta_i \, \left[ \,  \psi_{i+}^\dagger \psi_{i+} - \psi_{i-}^\dagger \psi_{i-}\,   \right],
\label{HY3}
\eeq
which is the analog of Eqs.~(\ref{Hsdw2}) and (\ref{HY2}). So the phase diagram of $\mathcal{H}_3$ will be analogous
to that for $\mathcal{H}_1$ in Fig.~\ref{fig:xy}, with the chargons experiencing uniform N\'eel order given by
Eq.~(\ref{HY3}) in phase C.

One important difference between the cases with $\mathbb{Z}_2$ and U(1) gauge theory is that pure
U(1) gauge theory is always confining in two spatial dimensions due to the proliferation of monopoles. 
At $p=0$, this mechanism will lead to insulating states with valence bond solid order \cite{NRSS89,NRSS90,FSX11}.
At non-zero $p$, monopoles can be suppressed by Fermi surfaces of particles carrying U(1) gauge charges \cite{Hermele04,SSLee08},
and this mechanism can stablize a U(1)-ACL in the intermediate phase C. 
But it should be noted that there is a non-BCS pairing instability of such a metallic state \cite{MMSS14},
and after the Fermi surface has been gapped by pairing, monopole-induced confinement will reappear.

\subsection{Stripe order}
\label{sec:stripe}

Finally, we consider the case of collinear spin order at wavevectors not equal to $(\pi, \pi)$. By symmetry, such spin order
is accompanied by charge density wave order at twice the wavevector \cite{ZKE98}. Considering the case of uni-directional
stripes with wavevector ${\bm K}_x$ for simplicity, the ordering is described by Eq.~(\ref{norder}) with 
\beq
\Phi_{x \ell} = e^{i \phi} \, n_{\ell}, \label{stripe1}
\eeq
where $n_{\ell}$ is a real vector obeying Eq.~(\ref{n21}). The order parameter for a charge density wave at
wavevector $2 {\bm K}_x$ is $e^{2i \phi}$.

Following the discussion in the previous subsections, we now have to examine fractionalizations of the stripe order parameter
in Eq.~(\ref{stripe1}). This question has been considered in a 
number of previous works \cite{Zaanen01,Zaanen02,ZDS02,SSTM02,Mross12,Mross12a}, all of which used a $\mathbb{Z}_2$ gauge
theory to fractionalize $\Phi_{x \ell}$ into its charge and spin components, represented by $e^{i \phi}$ and $n_{\ell}$ respectively;
such a fractionalization is invariant under
\beq
e^{i \phi_i} \rightarrow s_i e^{i \phi_i} \quad , \quad n_\ell \rightarrow s_i n_\ell
\eeq
with $s_i = \pm 1$ the $\mathbb{Z}_2$ gauge transformation. However, this fractionalization is not suitable for our purposes
because it does not involve spinor variables, and so cannot yield Fermi surface reconstruction in the phase with topological order.
So we examine combining the above fractionalization with the spinor decomposition of $n_\ell$ in Section~\ref{sec:neel}:
\beq
\Phi_{x \ell} = e^{i \phi}\, z_\alpha^\ast \sigma^\ell_{\alpha\beta} z_\beta . \label{stripe2}
\eeq
Then, in terms of the chargon variables in Eq.~(\ref{czpsi}), the Yukawa term coupling the order parameter to the
fermions (analogous to Eqs.~(\ref{Hsdw2}), (\ref{HY2}) and (\ref{HY3})) becomes
\beq
H_{Y} = - \lambda \sum_i 2 \cos \left(\phi + {\bm K}_x \cdot {\bm r}_i \right) \, \left[ \,  \psi_{i+}^\dagger \psi_{i+} - \psi_{i-}^\dagger \psi_{i-}\,   
\right].
\label{HY4}
\eeq
This expression makes it clear that the chargons move in the presence of a non-fluctuating potential only in a state with
long-range charge density wave order with $\langle e^{i \phi} \rangle \neq 0$. In such a situation, the theory of 
SDW order in the stripe model reduces \cite{SS09} to the U(1) gauge theory  
already considered in Section~\ref{sec:neel}. 
So Fermi surface reconstruction in a state with topological order requires charge density wave order in the stripe model.

\section{Conclusions}
\label{sec:conc}

This paper has presented an alternative approach to symmetry breaking and topological order in doped
Mott insulators. Instead of the conventional focus on electron fractionalization, we set up a formalism based
upon order parameter fractionalization. So our main Hamiltonian for the case of a XY SDW 
order parameter in Eq.~(\ref{H1}) involved a fractionalized `square root' of the SDW order parameter, but retained
the unfractionalized bare electron operator. Starting from Eq.~(\ref{H1}), we obtained the phase
diagram in Fig.~\ref{fig:xy}, containing states that had previously been obtained from the more common electron
fractionalization route. The advantage of our formalism is that offers a focus on just the phases observed
in experiments, while being very economical in using extraneous degrees of freedom which have to
be projected out. Furthermore, generalizations of the models
in Eqs.~(\ref{H1}) and (\ref{H1p}) are amenable to sign-problem-free quantum Monte
Carlo simulation by the methods of Refs.~\onlinecite{BMS12,SGTB15,LWYL15}, which can also study the connection to 
superconductivity.

We began with the LGW-Hertz theory for the onset of SDW order in metals:
this exhibits phases A and B in Fig.~\ref{fig:xy}, the Fermi liquid with large Fermi surface, and the SDW metal
with small pocket Fermi surfaces. Both phases have well-defined electronic quasiparticles and their Fermi surfaces sizes
obey the conventional Luttinger theorem. The phase transition between A and B has also been extensively 
studied \cite{Millis93,abanov00,ACS03,MMSS10b,LeeStrack13,SurLee15,PatelStrack15,Strack16}.
(In the limit of zero doping, $p=0$, phase A remains a Fermi liquid, while phase B becomes a Slater insulator
with long-range antiferromagnetic order.)

We argued that the LGW-Hertz theory could be modified to Eq.~(\ref{H1}) by introducing a $\mathbb{Z}_2$ gauge field,
and this allowed a phase transition in which the destruction of SDW order did not coincide with appearance of a
large Fermi surface: this led to phase C with pocket Fermi surfaces (of chargons and/or electrons)
and no SDW order. (In the limit $p=0$, phase C becomes a Mott insulator with $\mathbb{Z}_2$ topological order.)
For the case of easy-plane SDW order, we showed that the transition from phase B to phase
C was associated with the proliferation of doubled vortices. The universality class of the B-C transition has been
identified in earlier work \cite{TGTS10,RKK08b} as a relativistic 2+1 dimensional O(2)* field theory for the 
easy plane case (O(4)* for the Heisenberg case) \cite{SWSS16}.

Phase C (or more properly, its SU(2) spin rotation analogs in Section~\ref{sec:su2}) is proposed
as the pseudogap state of the hole-doped cuprates, present in between the phase B at low $p$,
and phase A above optimal doping. Other less-correlated high temperature superconductors (such as the pnictides)
are proposed to go directly from phase B to phase A. We maintain that this simple connection between different families of superconductors supports our model, and the unified phase diagram in Fig.~\ref{fig:xy}.

The main problem left open in our analysis is the nature of the transition from phase C to phase A.
This is a candidate transition for the physics near optimally hole-doped cuprate superconductors. 
The simple model in Eq.~(\ref{H1}) contains such a transition, but does not easily yield a continuum theory for
the quantum criticality. Phase C is in a deconfined phase of a $\mathbb{Z}_2$ gauge theory, while phase A is in
a confined phase. However, the transition between them is not just a $\mathbb{Z}_2$ confinement transition:
the $\mathbb{Z}_2$ gauge theory also changes from an ``odd'' gauge theory \cite{RJSS91,MVSS99,TSMPAF00,MSF02} 
(with $e^{2 \pi N_i } = -1$) in phase C
to an ``even'' gauge theory (with $e^{2 \pi N_i} = 1$) in phase A. 
Continuum formulations of confinement transitions in
$\mathbb{Z}_2$ gauge theories in the presence of gauge-charged matter require 
duality transforms to vison fields via mutual Chern-Simons terms \cite{CXSS09}, which we have not discussed here.
It is possible that such an analysis of the criticality will eventually lead to the deconfined SU(2) gauge theory for the C-A transition
proposed in Refs.~\onlinecite{SS09,DCSS15b,DCSS15,SSDC16}.

\subsection*{Acknowledgements}

We thank D. Chowdhury, M. Metlitski, T. Senthil, and A. Vishwanath for useful discussions.
The research was supported by the NSF under Grant DMR-1360789, by the ISF under Grant 1291/12, by the BSF under Grant 2014209, 
and by a Marie Curie CIG grant.
Research at Perimeter Institute is supported by the Government of Canada through Industry Canada and by the Province of Ontario 
through the Ministry of Research and Innovation. SS also acknowledges support from Cenovus Energy at Perimeter Institute.

\appendix

\section{Additional phases} 
\label{app:phases}

In our phase diagram in Fig.~\ref{fig:xy} for the easy-plane 
Hamiltonian $\mathcal{H}_1$ in Eq.~(\ref{H1}), we have 2 phases with
no broken symmetry, phases A and C. In phase C we are in the deconfined phase of the $\mathbb{Z}_2$ gauge theory, 
and the spinon number obeys $e^{2 i \pi N_i} = -1$
on each site. In contrast, in phase A we are in the confined phase of the $\mathbb{Z}_2$ gauge theory, and the spinon number 
obeys $e^{2 i \pi N_i} = 1$ on each site. There is no fundamental reason for these assignments of spinon number, and we
can also imagine additional phases with the opposite assignment; such phases are not shown in Fig.~\ref{fig:xy}.

A $\mathbb{Z}_2$ deconfined phase with $e^{2 i \pi N_i} = 1$ would have Fermi surfaces of chargons and/or electrons
of total size $1+p$, by the flux piercing arguments in Refs.~\onlinecite{TSMVSS04,APAV04,SSDC16}.
However, such a phase is energetically disfavored at large $\lambda$, and so not suitable for the physics of
the Mott-Hubbard systems under consideration here. 

More relevant is a $\mathbb{Z}_2$ confined phase with $e^{2 i \pi N_i} = -1$. 
This must have valence bond solid (VBS) order, as established in 
early work \cite{RJSS91,MVSS99,TSMPAF00}. More recent work has shown \cite{PCAS16} that the VBS
order can have a variety of complex spatial configurations, depending upon the nature of frustrating
interactions.
The Fermi surface is expected to be small by the flux piercing arguments, but with the doubling of the
unit cell by the VBS order, there is no fundamental distinction between small and large Fermi surfaces. 
Such a confining phase with VBS order is a possibility in Mott-Hubbard models with frustrated exchange
 interactions \cite{RKK07}, but is not shown in Fig.~\ref{fig:xy}.

\section{Single-site eigenstates with SU(2) symmetry} 

\subsection{O(4) model: Spiral order}
\label{app:OnSiteSpiral}
In the limit $K \rightarrow \infty$, the fluctuations of the $\mathbb{Z}_2$ gauge field are frozen, and we choose the gauge $\mu^{z}_{ij} = 1$. Then the model in Eq.~(\ref{H2}) reduces to:
\bea
\mathcal{H}_2^\prime &=& - \sum_{i,j} \left(t_{ij} + \mu \delta_{ij} \right) c_{i \alpha}^\dagger c_{j \alpha} - \lambda\sum_i \left[ \,  \varepsilon_{\alpha\gamma} z_{i\gamma} \sigma^{\ell}_{\alpha\beta} z_{i\beta} \, e^{ i {\bm K}_x 
\cdot{\bm r}_i } + \mbox{c.c.}  \right]
\, c_{i\alpha}^\dagger {\sigma}^\ell_{\alpha\beta} c_{i\beta}  \nn
 && -  \sum_{i< j} J_{ij} \, \left( z_{i\alpha}^\ast z_{j \alpha} + \mbox{c.c.} \right)
+ \Delta \sum_i \vec{L}_i^2  
\eea
Analogous to the $\mathbb{Z}_2$ case, the $J_{ij}$ term involves coupling to the fractionalized spinon-fields $z_{\alpha}$, which has $S_z$ quantized in units of $1/2$ on every site. We now further specialize to the strong coupling limit as defined by Eq.~(\ref{sclimit}), and show that the ground state at $p=0$ is again a Mott insulator with odd $\mathbb{Z}_2$ topological order. To do this, we define the on-site Hamiltonian $H_o$ by dropping the site index $i$ and letting $\xi_i = e^{i \bK \cdot \br_i}$: 
\bea
H_o &=&  -\mu \, c_{ \alpha}^\dagger c_{ \alpha} - \lambda \left[ \,  \varepsilon_{\alpha\gamma} z_{ \gamma} \sigma^{\ell}_{\alpha\beta} z_{ \beta} \, \xi_i + \mbox{c.c.}  \right]
\, c_{\alpha}^\dagger {\sigma}^\ell_{\alpha\beta} c_{\beta} + \Delta \sum_{\mu =1}^{6} \vec{L}_{\mu}^2  \nn
& = & -\mu \, c_{ \alpha}^\dagger c_{ \alpha}  -2 \lambda \left[ \left( \xi_i z_{\da}^2 - \xi_i^{*}(z_{\ua}^*)^2 \right) c^{\dagger}_{\da} c_{\ua} + \left( - \xi_i z_{\ua}^2 + \xi_i^{*} (z_{\da}^*)^2 \right) c^{\dagger}_{\ua} c_{\da} + (\xi_i z_{\da} z_{\ua} + \xi_i^* z_{\ua}^* z_{\da}^* ) ( c^{\dagger}_{\ua} c_{\ua} -  c^{\dagger}_{\da} c_{\da})\right] \nn 
&& + \Delta \sum_{\mu =1}^{6} \vec{L}_{\mu}^2
\eea
We start by looking at the eigenmodes of the O(4) rotor angular momenta $ \sum_{\mu=1}^{6} \vec{L}_{\mu}^2 $. These are given by the hyperspherical harmonics $Y^{n}_{l,m}$, which are a  complete set of eigenfunctions on the 3-sphere $S^{3}$ (generalizations of the spherical harmonics $Y^{l}_{m}$ on $S^{2}$).
\beq
\left( \sum_{\mu} \vec{L}_{\mu}^2 \right) Y^{n}_{l,m}  = n(n+2) Y^{n}_{l,m}, ~~~ n \in \{0,1,2,...\}
\eeq
We can conveniently describe these eigenmodes in the toroidal coordinates \cite{lehoucq2003}, which we define as follows:
\beq
( z_{\ua}, z_{\da} ) = (\text{cos}(\beta) e^{- i \theta} , \text{ sin}(\beta) e^{- i \phi}), ~~~ \text{ where } 0 \leq \beta \leq \pi/2, ~~ 0 \leq \theta, \phi < 2 \pi 
\eeq
In these coordinates, the hyperspherical harmonics are given by (here we choose a slightly different basis compared to Ref.~\onlinecite{lehoucq2003} for later computational convenience)
\bea
Y^{n}_{l,m} &=& \mathcal{N}^{n}_{l,m} \frac{e^{i l \theta}}{\sqrt{2 \pi}} \frac{e^{i m \phi}}{\sqrt{2 \pi}} \text{ cos}^{| l |}(\beta) \text{ sin}^{|m|}(\beta) P_{d}^{(|l|,|m|)}(\text{cos}(2 \beta)), ~~~ d = \frac{n - (|l| + |m|)}{2} \in \mathbb{Z}, |l| + |m| \leq n \nn
\eea
where $P_{d}^{(|l|,|m|)}(u)$ are the Jacobi polynomials and $\mathcal{N}^{n}_{l,m}$ are appropriate normalization constants, which we provide explicitly below for completeness:
\bea
P_{d}^{(|l|,|m|)}(u) & = & \frac{1}{2^{d}} \sum_{i=0}^{d} \binom{|m| + d}{i} \binom{|l| + d}{d-i} (u + 1)^{i}(u - 1)^{d-i} \nn
\mathcal{N}^{n}_{l,m} & = & \sqrt{\frac{2 (n + 1) d! (|l| + |m| + d)!}{(|l|+ d)! (|m| + d)!}}
\eea
Using the above coordinates, we list the important low-lying eigenstates.

(i) \underline{Mott insulator} \\
In analogy with the easy-axis case, we look for the lowest energy state with a single electron per site in the $n=1$ subspace of the hyperspherical harmonics. We choose the following basis, labeling the states as $c^{\dagger}_{\sigma}Y^{1}_{l,m}\ket{0}$:
\beq
\{ c^{\dagger}_{\ua}Y^{1}_{1,0}\ket{0}, c^{\dagger}_{\ua}Y^{1}_{-1,0}\ket{0}, c^{\dagger}_{\ua}Y^{1}_{0,1}\ket{0}, c^{\dagger}_{\ua}Y^{1}_{0,-1}\ket{0},  c^{\dagger}_{\da}Y^{1}_{1,0}\ket{0},c^{\dagger}_{\da}Y^{1}_{-1,0}\ket{0}, c^{\dagger}_{\da}Y^{1}_{0,1}\ket{0}, c^{\dagger}_{\da}Y^{1}_{0,-1}\ket{0} \}  
\label{MottBasis}
\eeq
In this subspace, we have:
\bea
H_o = (-\mu + 3 \Delta) \mathbb{I}_{8 \times 8}  - \frac{2 \lambda}{3} \begin{pmatrix}
0 & 0 & 0 & \xi_i^* & 0 & 0 & 0 & 0 \\
0 & 0 & \xi_i & 0 & -2 \xi_i & 0 & 0 & 0 \\
0 & \xi_i^* & 0 & 0 & 0 & 0 & 0 & 2 \xi_i^* \\
\xi_i & 0 & 0 & 0 & 0 & 0 & 0 & 0 \\
0 & - 2 \xi_i^* & 0 & 0 & 0 & 0 & 0 & - \xi_i^* \\
0 & 0 & 0 & 0 & 0 & 0 & - \xi_i & 0 \\
0 & 0 & 0 & 0 & 0 & -\xi_i^* & 0 & 0 \\
0 & 0 & 2 \xi_i  & 0 & -\xi_i & 0 & 0 & 0 \\
\end{pmatrix}
\eea
The lowest-energy Mott insulating state $\ket{G}$, with energy $ E_G = - \mu + 3 \Delta - 2 \lambda$, is given by:
\bea
\ket{G} &=& \frac{1}{2}\left( c^{\dagger}_{\ua}Y^{1}_{-1,0}+ \xi_i^* c^{\dagger}_{\ua}Y^{1}_{0,1} - \xi_i^* c^{\dagger}_{\da}Y^{1}_{1,0} +  c^{\dagger}_{\da}Y^{1}_{0,-1} \right)\ket{0} \nn 
& = & \frac{1}{2 \pi} \left[ \left( c^{\dagger}_{\ua} z_{\ua}  + c^{\dagger}_{\da}z_{\da} \right) - \xi_{i}^* \left( -c^{\dagger}_{\ua} z_{\da}^* + c^{\dagger}_{\da}z_{\ua}^* \right) \right]\ket{0}  \sim   \left( \psi^{\dagger}_{+}  - \xi_i^* \, \psi^{\dagger}_{-} \right)\ket{0} 
\eea 
where $\psi_{\pm}$ are the spinless fermionic chargons defined in Eq.~(\ref{czpsi}). The last representation makes it evident that the Mott insulating ground state does not carry any spin. However, it carries $\mathbb{Z}_2$ gauge charge, and therefore the $\mathbb{Z}_2$ gauge theory is odd. 

(ii) \underline{Spinons} \\
These are doubly degenerate states which have $S_z = \pm 1/2$, and a $\mathbb{Z}_2$ gauge charge relative to the Mott insulator, but no electromagnetic charge. We consider only the $\ket{\ua}$ spinon, the calculations for the $\ket{\da}$ spinon are idenical. Therefore, we choose the following basis of states which span the subspace with $S_z = 1/2$:
\beq
\{ c^{\dagger}_{\ua}Y^{0}_{0,0}\ket{0}, ~ c^{\dagger}_{\ua}Y^{2}_{0,0}\ket{0}, ~ c^{\dagger}_{\ua}Y^{2}_{1,1}\ket{0}, ~ c^{\dagger}_{\ua}Y^{2}_{-1,-1}\ket{0}, ~ c^{\dagger}_{\da}Y^{2}_{2,0}\ket{0}, ~ c^{\dagger}_{\da}Y^{2}_{0,-2}\ket{0}, ~ c^{\dagger}_{\da}Y^{2}_{1,-1}\ket{0} \}
\label{SpinonBasis}
\eeq
In this basis, we have:
\bea
H = -\mu \, \mathbb{I}_{7 \times 7} + \begin{pmatrix}
0 & 0 & -\sqrt{\frac{2}{3}} \lambda \xi_i & -\sqrt{\frac{2}{3}} \lambda \xi_i^* & \frac{2 \lambda \xi_i}{\sqrt{3}} &  -\frac{2 \lambda \xi_i^*}{\sqrt{3}} & 0 \\
0 & 8 \Delta  & 0 & 0 & \lambda \xi_i & \lambda \xi_i^{*} & 0 \\
-\sqrt{\frac{2}{3}} \lambda \xi_i^*  & 0 & 8 \Delta & 0 & 0 & 0 & -\lambda \xi_i^{*} \\
 -\sqrt{\frac{2}{3}} \lambda \xi_i & 0 & 0 & 8 \Delta & 0 & 0 & \lambda \xi_i \\
\frac{2 \lambda \xi_i^*}{\sqrt{3}} & \lambda \xi_i^{*} & 0 & 0 & 8 \Delta & 0 & \frac{\lambda \xi_i^*}{\sqrt{2}}  \\
-\frac{2 \lambda \xi_i}{\sqrt{3}} & \lambda \xi_i & 0 & 0 & 0 & 8 \Delta & \frac{\lambda \xi_i}{\sqrt{2}} \\
0 & 0 & - \lambda \xi_i &  \lambda \xi_i^* &  \frac{\lambda \xi_i}{\sqrt{2}} &  \frac{\lambda \xi_i^*}{\sqrt{2}} & 8 \Delta \\
\end{pmatrix}
\label{HSpinonSU2}
\eea

For all positive values of $\lambda$ and $\Delta$, we find that the energy of the lowest-lying spinon state is $E_s = - \mu +  4 \Delta - \sqrt{ 16 \Delta^2 + 4 \lambda^2} $. For the Mott insulator to have lower energy than the spinon, we require $E_G < E_s$, which translates to $\lambda > 15 \Delta/4$. The spin-gap of the Mott-insulator is given by:
\beq
\Delta_s = E_s - E_G = 2 \lambda + \Delta - \sqrt{16 \Delta^2 + 4 \lambda^2}
\eeq

(iii) \underline{Holon}\\
This is the empty state $\ket{0}$, with energy $E_{hn} = 0$. Relative to the Mott insulator, it has $S_z=0$, electromagnetic charge $+e$ and non-zero $\mathbb{Z}_2$ gauge charge. 

(iv) \underline{Doublon}\\
This is the state $c^{\dagger}_{\ua}c^{\dagger}_{\da}\ket{0}$, with energy $E_d = - 2 \mu$.  Relative to the Mott insulator, it has $S_z=0$, electromagnetic charge $-e$ and non-zero $\mathbb{Z}_2$ gauge charge.

(v) \underline{Holes} \\
These are the 4 degenerate states given by the $n=1$ hyperspherical harmonics, which can be represented as $Y^{1}_{l,m}\ket{0}$, for $\{l= \pm 1, m=0\}$ and $\{ l=0,m=\pm 1 \}$. Each state has energy given by $E_h =  3 \Delta$. They have electromagnetic charge $+e$ and $S_z = \pm 1/2$ relative to the Mott insulator, but no $\mathbb{Z}_2$ gauge charge. The energy difference between a pair of sites with hole+Mott insulator and a pair with holon+spinon is given by:
\beq
E_G + E_h - E_{hn} - E_s = 2 \Delta - 2 \lambda + \sqrt{ 16 \Delta^2 + 4 \lambda^2} > 0
\label{holeIns}
\eeq
Therefore, the hole is unstable to decay to a holon and a spinon in the strong-coupling limit.

(vi) \underline{Electrons} \\
These are again 4 degenerate states given by $c^{\dagger}_{\ua}c^{\dagger}_{\da}Y^{1}_{l,m}\ket{0}$ for $\{l= \pm 1, m=0\}$ and $\{ l=0,m=\pm 1 \}$. One can check that $H_{int}$ acting on any of these states gives zero, so only the diagonal terms matter and therefore, they have energy $E_e = - 2\mu + 3 \Delta$. They have electromagnetic charge $-e$ and $S_z = \pm 1/2$ relative to the Mott insulator, but no $\mathbb{Z}_2$ gauge charge. The energy difference between a pair of sites with electron+Mott-insulator and a pair with doublon+spinon is given by:
\beq
E_G + E_e - E_{dn} - E_s = 2 \Delta - 2 \lambda + \sqrt{ 16 \Delta^2 + 4 \lambda^2} > 0
\eeq
which is the same condition as Eq.~(\ref{holeIns}). Therefore, the electron is also unstable to decay to a doublon and a spinon in the strong-coupling limit. 

Finally, note that we can define a new spinless fermionic operator as a linear combination of $\psi_{i,\pm}$, given by:
\beq
\psi_i = \frac{1}{\sqrt{2}} \left( \psi_{i,+}  - \xi_i \, \psi_{i,-}  \right)
\eeq
$\psi$ is the holon creation operator, and the Mott insulator is a filled band of $\psi$, given by 
\beq
\ket{G} =  \prod_i \psi_i^\dagger \ket{0}
\eeq
We can also define bosonic spinon creation operators analogous to Eq.~(\ref{spinonOp}) which will create $S_z = \pm 1/2$ excitations over the Mott insulating ground state using the eigenstates of Eq.~(\ref{HSpinonSU2}). 

\subsection{O(4) model: N\'eel order}
\label{app:OnSiteNeel}

For the fractionalization defined by $n_{\ell} = z_\alpha^\ast \sigma^\ell_{\alpha\beta} z_\beta$, the constraint defined by Eq.~(\ref{n21}) can be re-written as $|z_\uparrow|^2 + |z_\downarrow|^2 = 1$. Therefore, the dynamics of the order parameter field are described by an O(4) model which is coupled to the $c$ fermions. We again consider the limits of $K \rightarrow \infty$ and strong-coupling as defined by Eq.~(\ref{sclimit}), resulting in the following on-site Hamiltonian $H_o$ (with $\eta_i = (-1)^{x_i + y_i}$):
 \bea
H_{o} &=& - \mu \, c_{ \alpha}^\dagger c_{ \alpha} - \lambda \eta_i \, z^{*}_{\alpha}\sigma^l_{\alpha \beta} z_{\beta} \, c_{\alpha}^\dagger {\sigma}^\ell_{\alpha\beta} c_{\beta} + \Delta \sum_{\mu} \vec{L}_{\mu}^2 \nn
& = & - \mu \, c_{ \alpha}^\dagger c_{ \alpha}  - \lambda \eta_i \left[ 2 z^*_{\ua}z_{\da} c^{\dagger}_{\da} c_{\ua} + 2 z^{*}_{\da}z_{\ua} c^{\dagger}_{\ua} c_{\da} + ( z^*_{\ua} z_{\ua} - z_{\da}^* z_{\da} ) ( c^{\dagger}_{\ua} c_{\ua} -  c^{\dagger}_{\da} c_{\da})\right] + \Delta \sum_{\mu} \vec{L}_{\mu}^2
\eea
The rest of the calculation exactly follows Appendix \ref{app:OnSiteSpiral}. Only the Mott-insulator and the spinons eigenstates are different, so we restrict the following description to these two kinds of eigenstates. 

(i) \underline{Mott insulator}:
In the basis described in Eq.~(\ref{MottBasis}) we have:
\beq
H_o = (-\mu + 3\Delta) \mathbb{I}_{8 \times 8}  - \frac{\lambda \eta_i}{3} \begin{pmatrix}
1 & 0 & 0 & 0 & 0 & 0 & 0 & 0 \\
0 & 1  & 0 & 0 & 0 & 0 & 0 & 2 \\
0 & 0 & -1 & 0 & 2 & 0 & 0 & 0 \\
0 & 0 & 0 & -1 & 0 & 0 & 0 & 0 \\
0 & 0 & 2 & 0 & -1 & 0 & 0 & 0 \\
0& 0 & 0 & 0 & 0 & -1 & 0 & 0 \\
0 & 0 & 0 & 0 & 0 & 0 & 1 & 0 \\
0 & 2 & 0 & 0 & 0& 0 & 0 & 1 \\
\end{pmatrix}
\eeq
The lowest energy state is given by:
\beq
\ket{G} = \begin{cases}
\frac{1}{\sqrt{2}}  \left(c^{\dagger}_{\ua}Y^{1}_{-1,0} + c^{\dagger}_{\da} Y^{1}_{0,-1}\right)\ket{0} = \frac{1}{2\pi} \left(c^{\dagger}_{\ua} z_{\ua} + c^{\dagger}_{\da} z_{\da}\right)\ket{0} \sim \psi_{+}^{\dagger}\ket{0}, \text{   for } \eta_i = 1, \nn
 \frac{1}{\sqrt{2}}  \left(-c^{\dagger}_{\ua}Y^{1}_{0,1} + c^{\dagger}_{\da} Y^{1}_{1,0}\right)\ket{0} =  \frac{1}{2\pi} \left(-c^{\dagger}_{\ua} z^*_{\da} + c^{\dagger}_{\da} z^*_{\ua}\right)\ket{0} \sim \psi_{-}^{\dagger}\ket{0}, \text{   for } \eta_i = -1 
\end{cases}
\eeq
Therefore, the Mott insulator can be written conveniently in terms of the spinless fermionic chargons as:
\beq
\ket{G} = \psi_{\eta_i}^{\dagger}\ket{0}, ~~~ E_G = - \mu + 3\Delta - \lambda
\eeq
In this representation, it is evident that $\ket{G}$ does not carry any spin but carries a non-zero $\mathbb{Z}_2$ gauge charge, and and therefore the $\mathbb{Z}_2$ gauge theory is odd. 

(ii) \underline{Spinons}:
These are doubly degenerate states which have $S_z = \pm 1/2$, and a $\mathbb{Z}_2$ gauge charge relative to the Mott insulator, but no electromagnetic charge. We can find the $\ket{\ua}$ spinon by using the $S_z = 1/2$ subspace defined in Eq.~(\ref{SpinonBasis}).
\beq
H_o = -\mu \mathbb{I}_{7 \times 7} + \begin{pmatrix}
0 & - \frac{\lambda \eta_i}{\sqrt{3}} & 0 & 0 & 0 & 0 & -\sqrt{\frac{2}{3}}\lambda \eta_i  \\
- \frac{\lambda \eta_i}{\sqrt{3}} & 8 \Delta  & 0 & 0 & 0 & 0  & 0 \\
0 & 0 &  8 \Delta & 0 & - \frac{\lambda \eta_i}{\sqrt{2}} & 0 & 0 \\
0 & 0 & 0 & 8 \Delta & 0 & - \frac{\lambda \eta_i}{\sqrt{2}} & 0 \\
0 & 0 & - \frac{\lambda \eta_i}{\sqrt{2}} & 0 &  8 \Delta + \frac{\lambda \eta_i}{2} & 0 & 0 \\
0& 0 & 0 & - \frac{\lambda \eta_i}{\sqrt{2}} & 0 & 8 \Delta - \frac{\lambda \eta_i}{2} & 0 \\
-\sqrt{\frac{2}{3}}\lambda \eta_i  & 0 & 0 & 0 & 0 & 0 & 8 \Delta \\
\end{pmatrix}
\eeq

The energy of the lowest-lying spinon state is $E_s = - \mu +  4 \Delta - \sqrt{ 16 \Delta^2 + \lambda^2} $. For the Mott insulator to have lower energy than the spinon, we require $E_G < E_s$, which translates to $\lambda > 15 \Delta/2$. The spin-gap of the Mott-insulator is given by:
\beq
\Delta_s = E_s - E_G =  \lambda + \Delta - \sqrt{16 \Delta^2 + \lambda^2}
\eeq

The descriptions of the (iii) \underline{holon}, (iv) \underline{doublon}, (v) \underline{holes} and (vi) \underline{electrons} are identical to Appendix \ref{app:OnSiteSpiral}. The only difference arises from the change in energy eigenvalues $E_G$ and $E_s$. This changes the energy gap between a pair of sites with hole+Mott insulator (electron + Mott insulator) and a pair with holon+spinon (doublon+spinon):
\beq
E_G + E_h - E_{hn} - E_s = E_G + E_e - E_{dn} - E_s = 2 \Delta -  \lambda + \sqrt{ 16 \Delta^2 +  \lambda^2} > 0
\eeq
As before, we observe that the hole (electron) is unstable to decay to a holon (doublon) and a spinon in the strong-coupling limit.

\bibliography{sdwfrac}

\begin{thebibliography}{72}%
\makeatletter
\providecommand \@ifxundefined [1]{%
 \@ifx{#1\undefined}
}%
\providecommand \@ifnum [1]{%
 \ifnum #1\expandafter \@firstoftwo
 \else \expandafter \@secondoftwo
 \fi
}%
\providecommand \@ifx [1]{%
 \ifx #1\expandafter \@firstoftwo
 \else \expandafter \@secondoftwo
 \fi
}%
\providecommand \natexlab [1]{#1}%
\providecommand \enquote  [1]{``#1''}%
\providecommand \bibnamefont  [1]{#1}%
\providecommand \bibfnamefont [1]{#1}%
\providecommand \citenamefont [1]{#1}%
\providecommand \href@noop [0]{\@secondoftwo}%
\providecommand \href [0]{\begingroup \@sanitize@url \@href}%
\providecommand \@href[1]{\@@startlink{#1}\@@href}%
\providecommand \@@href[1]{\endgroup#1\@@endlink}%
\providecommand \@sanitize@url [0]{\catcode `\\12\catcode `\$12\catcode
  `\&12\catcode `\#12\catcode `\^12\catcode `\_12\catcode `\%12\relax}%
\providecommand \@@startlink[1]{}%
\providecommand \@@endlink[0]{}%
\providecommand \url  [0]{\begingroup\@sanitize@url \@url }%
\providecommand \@url [1]{\endgroup\@href {#1}{\urlprefix }}%
\providecommand \urlprefix  [0]{URL }%
\providecommand \Eprint [0]{\href }%
\providecommand \doibase [0]{http://dx.doi.org/}%
\providecommand \selectlanguage [0]{\@gobble}%
\providecommand \bibinfo  [0]{\@secondoftwo}%
\providecommand \bibfield  [0]{\@secondoftwo}%
\providecommand \translation [1]{[#1]}%
\providecommand \BibitemOpen [0]{}%
\providecommand \bibitemStop [0]{}%
\providecommand \bibitemNoStop [0]{.\EOS\space}%
\providecommand \EOS [0]{\spacefactor3000\relax}%
\providecommand \BibitemShut  [1]{\csname bibitem#1\endcsname}%
\let\auto@bib@innerbib\@empty
%</preamble>
\bibitem [{\citenamefont {{Mirzaei}}\ \emph {et~al.}(2013)\citenamefont
  {{Mirzaei}}, \citenamefont {{Stricker}}, \citenamefont {{Hancock}},
  \citenamefont {{Berthod}}, \citenamefont {{Georges}}, \citenamefont {{van
  Heumen}}, \citenamefont {{Chan}}, \citenamefont {{Zhao}}, \citenamefont
  {{Li}}, \citenamefont {{Greven}}, \citenamefont {{Barisic}},\ and\
  \citenamefont {{van der Marel}}}]{Marel13}%
  \BibitemOpen
  \bibfield  {author} {\bibinfo {author} {\bibfnamefont {S.~I.}\ \bibnamefont
  {{Mirzaei}}}, \bibinfo {author} {\bibfnamefont {D.}~\bibnamefont
  {{Stricker}}}, \bibinfo {author} {\bibfnamefont {J.~N.}\ \bibnamefont
  {{Hancock}}}, \bibinfo {author} {\bibfnamefont {C.}~\bibnamefont
  {{Berthod}}}, \bibinfo {author} {\bibfnamefont {A.}~\bibnamefont
  {{Georges}}}, \bibinfo {author} {\bibfnamefont {E.}~\bibnamefont {{van
  Heumen}}}, \bibinfo {author} {\bibfnamefont {M.~K.}\ \bibnamefont {{Chan}}},
  \bibinfo {author} {\bibfnamefont {X.}~\bibnamefont {{Zhao}}}, \bibinfo
  {author} {\bibfnamefont {Y.}~\bibnamefont {{Li}}}, \bibinfo {author}
  {\bibfnamefont {M.}~\bibnamefont {{Greven}}}, \bibinfo {author}
  {\bibfnamefont {N.}~\bibnamefont {{Barisic}}}, \ and\ \bibinfo {author}
  {\bibfnamefont {D.}~\bibnamefont {{van der Marel}}},\ }\bibfield  {title}
  {\enquote {\bibinfo {title} {{Spectroscopic evidence for Fermi liquid-like
  energy and temperature dependence of the relaxation rate in the pseudogap
  phase of the cuprates}},}\ }\href {\doibase 10.1073/pnas.1218846110}
  {\bibfield  {journal} {\bibinfo  {journal} {Proc. Nat. Acad. Sci.}\ }\textbf
  {\bibinfo {volume} {110}},\ \bibinfo {pages} {5774} (\bibinfo {year}
  {2013})},\ \Eprint {http://arxiv.org/abs/1207.6704} {arXiv:1207.6704
  [cond-mat.supr-con]} \BibitemShut {NoStop}%
\bibitem [{\citenamefont {{He}}\ \emph {et~al.}(2014)\citenamefont {{He}},
  \citenamefont {{Yin}}, \citenamefont {{Zech}}, \citenamefont
  {{Soumyanarayanan}}, \citenamefont {{Yee}}, \citenamefont {{Williams}},
  \citenamefont {{Boyer}}, \citenamefont {{Chatterjee}}, \citenamefont
  {{Wise}}, \citenamefont {{Zeljkovic}}, \citenamefont {{Kondo}}, \citenamefont
  {{Takeuchi}}, \citenamefont {{Ikuta}}, \citenamefont {{Mistark}},
  \citenamefont {{Markiewicz}}, \citenamefont {{Bansil}}, \citenamefont
  {{Sachdev}}, \citenamefont {{Hudson}},\ and\ \citenamefont
  {{Hoffman}}}]{YHe13}%
  \BibitemOpen
  \bibfield  {author} {\bibinfo {author} {\bibfnamefont {Y.}~\bibnamefont
  {{He}}}, \bibinfo {author} {\bibfnamefont {Y.}~\bibnamefont {{Yin}}},
  \bibinfo {author} {\bibfnamefont {M.}~\bibnamefont {{Zech}}}, \bibinfo
  {author} {\bibfnamefont {A.}~\bibnamefont {{Soumyanarayanan}}}, \bibinfo
  {author} {\bibfnamefont {M.~M.}\ \bibnamefont {{Yee}}}, \bibinfo {author}
  {\bibfnamefont {T.}~\bibnamefont {{Williams}}}, \bibinfo {author}
  {\bibfnamefont {M.~C.}\ \bibnamefont {{Boyer}}}, \bibinfo {author}
  {\bibfnamefont {K.}~\bibnamefont {{Chatterjee}}}, \bibinfo {author}
  {\bibfnamefont {W.~D.}\ \bibnamefont {{Wise}}}, \bibinfo {author}
  {\bibfnamefont {I.}~\bibnamefont {{Zeljkovic}}}, \bibinfo {author}
  {\bibfnamefont {T.}~\bibnamefont {{Kondo}}}, \bibinfo {author} {\bibfnamefont
  {T.}~\bibnamefont {{Takeuchi}}}, \bibinfo {author} {\bibfnamefont
  {H.}~\bibnamefont {{Ikuta}}}, \bibinfo {author} {\bibfnamefont
  {P.}~\bibnamefont {{Mistark}}}, \bibinfo {author} {\bibfnamefont {R.~S.}\
  \bibnamefont {{Markiewicz}}}, \bibinfo {author} {\bibfnamefont
  {A.}~\bibnamefont {{Bansil}}}, \bibinfo {author} {\bibfnamefont
  {S.}~\bibnamefont {{Sachdev}}}, \bibinfo {author} {\bibfnamefont {E.~W.}\
  \bibnamefont {{Hudson}}}, \ and\ \bibinfo {author} {\bibfnamefont {J.~E.}\
  \bibnamefont {{Hoffman}}},\ }\bibfield  {title} {\enquote {\bibinfo {title}
  {{Fermi Surface and Pseudogap Evolution in a Cuprate Superconductor}},}\
  }\href {\doibase 10.1126/science.1248221} {\bibfield  {journal} {\bibinfo
  {journal} {Science}\ }\textbf {\bibinfo {volume} {344}},\ \bibinfo {pages}
  {608} (\bibinfo {year} {2014})},\ \Eprint {http://arxiv.org/abs/1305.2778}
  {arXiv:1305.2778 [cond-mat.supr-con]} \BibitemShut {NoStop}%
\bibitem [{\citenamefont {{Fujita}}\ \emph {et~al.}(2014)\citenamefont
  {{Fujita}}, \citenamefont {{Kim}}, \citenamefont {{Lee}}, \citenamefont
  {{Lee}}, \citenamefont {{Hamidian}}, \citenamefont {{Firmo}}, \citenamefont
  {{Mukhopadhyay}}, \citenamefont {{Eisaki}}, \citenamefont {{Uchida}},
  \citenamefont {{Lawler}}, \citenamefont {{Kim}},\ and\ \citenamefont
  {{Davis}}}]{Fujita14a}%
  \BibitemOpen
  \bibfield  {author} {\bibinfo {author} {\bibfnamefont {K.}~\bibnamefont
  {{Fujita}}}, \bibinfo {author} {\bibfnamefont {C.~K.}\ \bibnamefont {{Kim}}},
  \bibinfo {author} {\bibfnamefont {I.}~\bibnamefont {{Lee}}}, \bibinfo
  {author} {\bibfnamefont {J.}~\bibnamefont {{Lee}}}, \bibinfo {author}
  {\bibfnamefont {M.~H.}\ \bibnamefont {{Hamidian}}}, \bibinfo {author}
  {\bibfnamefont {I.~A.}\ \bibnamefont {{Firmo}}}, \bibinfo {author}
  {\bibfnamefont {S.}~\bibnamefont {{Mukhopadhyay}}}, \bibinfo {author}
  {\bibfnamefont {H.}~\bibnamefont {{Eisaki}}}, \bibinfo {author}
  {\bibfnamefont {S.}~\bibnamefont {{Uchida}}}, \bibinfo {author}
  {\bibfnamefont {M.~J.}\ \bibnamefont {{Lawler}}}, \bibinfo {author}
  {\bibfnamefont {E.-A.}\ \bibnamefont {{Kim}}}, \ and\ \bibinfo {author}
  {\bibfnamefont {J.~C.}\ \bibnamefont {{Davis}}},\ }\bibfield  {title}
  {\enquote {\bibinfo {title} {{Simultaneous Transitions in Cuprate
  Momentum-Space Topology and Electronic Symmetry Breaking}},}\ }\href
  {\doibase 10.1126/science.1248783} {\bibfield  {journal} {\bibinfo  {journal}
  {Science}\ }\textbf {\bibinfo {volume} {344}},\ \bibinfo {pages} {612}
  (\bibinfo {year} {2014})},\ \Eprint {http://arxiv.org/abs/1403.7788}
  {arXiv:1403.7788 [cond-mat.supr-con]} \BibitemShut {NoStop}%
\bibitem [{\citenamefont {{Badoux}}\ \emph {et~al.}(2016)\citenamefont
  {{Badoux}}, \citenamefont {{Tabis}}, \citenamefont {{Lalibert{\'e}}},
  \citenamefont {{Grissonnanche}}, \citenamefont {{Vignolle}}, \citenamefont
  {{Vignolles}}, \citenamefont {{B{\'e}ard}}, \citenamefont {{Bonn}},
  \citenamefont {{Hardy}}, \citenamefont {{Liang}}, \citenamefont
  {{Doiron-Leyraud}}, \citenamefont {{Taillefer}},\ and\ \citenamefont
  {{Proust}}}]{LTCP15}%
  \BibitemOpen
  \bibfield  {author} {\bibinfo {author} {\bibfnamefont {S.}~\bibnamefont
  {{Badoux}}}, \bibinfo {author} {\bibfnamefont {W.}~\bibnamefont {{Tabis}}},
  \bibinfo {author} {\bibfnamefont {F.}~\bibnamefont {{Lalibert{\'e}}}},
  \bibinfo {author} {\bibfnamefont {G.}~\bibnamefont {{Grissonnanche}}},
  \bibinfo {author} {\bibfnamefont {B.}~\bibnamefont {{Vignolle}}}, \bibinfo
  {author} {\bibfnamefont {D.}~\bibnamefont {{Vignolles}}}, \bibinfo {author}
  {\bibfnamefont {J.}~\bibnamefont {{B{\'e}ard}}}, \bibinfo {author}
  {\bibfnamefont {D.~A.}\ \bibnamefont {{Bonn}}}, \bibinfo {author}
  {\bibfnamefont {W.~N.}\ \bibnamefont {{Hardy}}}, \bibinfo {author}
  {\bibfnamefont {R.}~\bibnamefont {{Liang}}}, \bibinfo {author} {\bibfnamefont
  {N.}~\bibnamefont {{Doiron-Leyraud}}}, \bibinfo {author} {\bibfnamefont
  {L.}~\bibnamefont {{Taillefer}}}, \ and\ \bibinfo {author} {\bibfnamefont
  {C.}~\bibnamefont {{Proust}}},\ }\bibfield  {title} {\enquote {\bibinfo
  {title} {{Change of carrier density at the pseudogap critical point of a
  cuprate superconductor}},}\ }\href {\doibase 10.1038/nature16983} {\bibfield
  {journal} {\bibinfo  {journal} {Nature}\ }\textbf {\bibinfo {volume} {531}},\
  \bibinfo {pages} {210} (\bibinfo {year} {2016})},\ \Eprint
  {http://arxiv.org/abs/1511.08162} {arXiv:1511.08162 [cond-mat.supr-con]}
  \BibitemShut {NoStop}%
\bibitem [{\citenamefont {{Chakravarty}}\ \emph {et~al.}(2002)\citenamefont
  {{Chakravarty}}, \citenamefont {{Nayak}}, \citenamefont {{Tewari}},\ and\
  \citenamefont {{Yang}}}]{SC03}%
  \BibitemOpen
  \bibfield  {author} {\bibinfo {author} {\bibfnamefont {S.}~\bibnamefont
  {{Chakravarty}}}, \bibinfo {author} {\bibfnamefont {C.}~\bibnamefont
  {{Nayak}}}, \bibinfo {author} {\bibfnamefont {S.}~\bibnamefont {{Tewari}}}, \
  and\ \bibinfo {author} {\bibfnamefont {X.}~\bibnamefont {{Yang}}},\
  }\bibfield  {title} {\enquote {\bibinfo {title} {{Sharp Signature of a
  $d_{x^2-y^2}$ Quantum Critical Point in the Hall Coefficient of Cuprate
  Superconductors}},}\ }\href {\doibase 10.1103/PhysRevLett.89.277003}
  {\bibfield  {journal} {\bibinfo  {journal} {Physical Review Letters}\
  }\textbf {\bibinfo {volume} {89}},\ \bibinfo {eid} {277003} (\bibinfo {year}
  {2002})},\ \Eprint {http://arxiv.org/abs/cond-mat/0201228} {cond-mat/0201228}
  \BibitemShut {NoStop}%
\bibitem [{\citenamefont {{Motoyama}}\ \emph {et~al.}(2007)\citenamefont
  {{Motoyama}}, \citenamefont {{Yu}}, \citenamefont {{Vishik}}, \citenamefont
  {{Vajk}}, \citenamefont {{Mang}},\ and\ \citenamefont {{Greven}}}]{Greven07}%
  \BibitemOpen
  \bibfield  {author} {\bibinfo {author} {\bibfnamefont {E.~M.}\ \bibnamefont
  {{Motoyama}}}, \bibinfo {author} {\bibfnamefont {G.}~\bibnamefont {{Yu}}},
  \bibinfo {author} {\bibfnamefont {I.~M.}\ \bibnamefont {{Vishik}}}, \bibinfo
  {author} {\bibfnamefont {O.~P.}\ \bibnamefont {{Vajk}}}, \bibinfo {author}
  {\bibfnamefont {P.~K.}\ \bibnamefont {{Mang}}}, \ and\ \bibinfo {author}
  {\bibfnamefont {M.}~\bibnamefont {{Greven}}},\ }\bibfield  {title} {\enquote
  {\bibinfo {title} {{Spin correlations in the electron-doped
  high-transition-temperature superconductor Nd$_{2-x}$Ce$_{x}$CuO$_{4\pm
  \delta}$}},}\ }\href {\doibase 10.1038/nature05437} {\bibfield  {journal}
  {\bibinfo  {journal} {Nature}\ }\textbf {\bibinfo {volume} {445}},\ \bibinfo
  {pages} {186} (\bibinfo {year} {2007})},\ \Eprint
  {http://arxiv.org/abs/cond-mat/0609386} {cond-mat/0609386} \BibitemShut
  {NoStop}%
\bibitem [{\citenamefont {{Senthil}}\ \emph {et~al.}(2003)\citenamefont
  {{Senthil}}, \citenamefont {{Sachdev}},\ and\ \citenamefont
  {{Vojta}}}]{TSSSMV03}%
  \BibitemOpen
  \bibfield  {author} {\bibinfo {author} {\bibfnamefont {T.}~\bibnamefont
  {{Senthil}}}, \bibinfo {author} {\bibfnamefont {S.}~\bibnamefont
  {{Sachdev}}}, \ and\ \bibinfo {author} {\bibfnamefont {M.}~\bibnamefont
  {{Vojta}}},\ }\bibfield  {title} {\enquote {\bibinfo {title} {{Fractionalized
  Fermi Liquids}},}\ }\href {\doibase 10.1103/PhysRevLett.90.216403} {\bibfield
   {journal} {\bibinfo  {journal} {Phys. Rev. Lett.}\ }\textbf {\bibinfo
  {volume} {90}},\ \bibinfo {eid} {216403} (\bibinfo {year} {2003})},\ \Eprint
  {http://arxiv.org/abs/cond-mat/0209144} {cond-mat/0209144} \BibitemShut
  {NoStop}%
\bibitem [{\citenamefont {{Senthil}}\ \emph {et~al.}(2004)\citenamefont
  {{Senthil}}, \citenamefont {{Vojta}},\ and\ \citenamefont
  {{Sachdev}}}]{TSMVSS04}%
  \BibitemOpen
  \bibfield  {author} {\bibinfo {author} {\bibfnamefont {T.}~\bibnamefont
  {{Senthil}}}, \bibinfo {author} {\bibfnamefont {M.}~\bibnamefont {{Vojta}}},
  \ and\ \bibinfo {author} {\bibfnamefont {S.}~\bibnamefont {{Sachdev}}},\
  }\bibfield  {title} {\enquote {\bibinfo {title} {{Weak magnetism and
  non-Fermi liquids near heavy-fermion critical points}},}\ }\href {\doibase
  10.1103/PhysRevB.69.035111} {\bibfield  {journal} {\bibinfo  {journal} {Phys.
  Rev. B}\ }\textbf {\bibinfo {volume} {69}},\ \bibinfo {eid} {035111}
  (\bibinfo {year} {2004})},\ \Eprint {http://arxiv.org/abs/cond-mat/0305193}
  {cond-mat/0305193} \BibitemShut {NoStop}%
\bibitem [{\citenamefont {{Paramekanti}}\ and\ \citenamefont
  {{Vishwanath}}(2004)}]{APAV04}%
  \BibitemOpen
  \bibfield  {author} {\bibinfo {author} {\bibfnamefont {A.}~\bibnamefont
  {{Paramekanti}}}\ and\ \bibinfo {author} {\bibfnamefont {A.}~\bibnamefont
  {{Vishwanath}}},\ }\bibfield  {title} {\enquote {\bibinfo {title} {{Extending
  Luttinger's theorem to $\mathbb{Z}_{2}$ fractionalized phases of matter}},}\
  }\href {\doibase 10.1103/PhysRevB.70.245118} {\bibfield  {journal} {\bibinfo
  {journal} {Phys. Rev. B}\ }\textbf {\bibinfo {volume} {70}},\ \bibinfo {eid}
  {245118} (\bibinfo {year} {2004})},\ \Eprint
  {http://arxiv.org/abs/cond-mat/0406619} {cond-mat/0406619} \BibitemShut
  {NoStop}%
\bibitem [{\citenamefont {{Sachdev}}\ \emph {et~al.}(2009)\citenamefont
  {{Sachdev}}, \citenamefont {{Metlitski}}, \citenamefont {{Qi}},\ and\
  \citenamefont {{Xu}}}]{SS09}%
  \BibitemOpen
  \bibfield  {author} {\bibinfo {author} {\bibfnamefont {S.}~\bibnamefont
  {{Sachdev}}}, \bibinfo {author} {\bibfnamefont {M.~A.}\ \bibnamefont
  {{Metlitski}}}, \bibinfo {author} {\bibfnamefont {Y.}~\bibnamefont {{Qi}}}, \
  and\ \bibinfo {author} {\bibfnamefont {C.}~\bibnamefont {{Xu}}},\ }\bibfield
  {title} {\enquote {\bibinfo {title} {{Fluctuating spin density waves in
  metals}},}\ }\href {\doibase 10.1103/PhysRevB.80.155129} {\bibfield
  {journal} {\bibinfo  {journal} {Phys. Rev. B}\ }\textbf {\bibinfo {volume}
  {80}},\ \bibinfo {eid} {155129} (\bibinfo {year} {2009})},\ \Eprint
  {http://arxiv.org/abs/0907.3732} {arXiv:0907.3732 [cond-mat.str-el]}
  \BibitemShut {NoStop}%
\bibitem [{\citenamefont {{Chowdhury}}\ and\ \citenamefont
  {{Sachdev}}(2015{\natexlab{a}})}]{DCSS15b}%
  \BibitemOpen
  \bibfield  {author} {\bibinfo {author} {\bibfnamefont {D.}~\bibnamefont
  {{Chowdhury}}}\ and\ \bibinfo {author} {\bibfnamefont {S.}~\bibnamefont
  {{Sachdev}}},\ }\bibfield  {title} {\enquote {\bibinfo {title} {{Higgs
  criticality in a two-dimensional metal}},}\ }\href {\doibase
  10.1103/PhysRevB.91.115123} {\bibfield  {journal} {\bibinfo  {journal} {Phys.
  Rev. B}\ }\textbf {\bibinfo {volume} {91}},\ \bibinfo {eid} {115123}
  (\bibinfo {year} {2015}{\natexlab{a}})},\ \Eprint
  {http://arxiv.org/abs/1412.1086} {arXiv:1412.1086 [cond-mat.str-el]}
  \BibitemShut {NoStop}%
\bibitem [{\citenamefont {{Chowdhury}}\ and\ \citenamefont
  {{Sachdev}}(2015{\natexlab{b}})}]{DCSS15}%
  \BibitemOpen
  \bibfield  {author} {\bibinfo {author} {\bibfnamefont {D.}~\bibnamefont
  {{Chowdhury}}}\ and\ \bibinfo {author} {\bibfnamefont {S.}~\bibnamefont
  {{Sachdev}}},\ }\bibfield  {title} {\enquote {\bibinfo {title} {{The Enigma
  of the Pseudogap Phase of the Cuprate Superconductors}},}\ }in\ \href
  {\doibase 10.1142/9789814704090_0001} {\emph {\bibinfo {booktitle} {Quantum
  criticality in condensed matter}}},\ \bibinfo {series and number} {50th
  Karpacz Winter School of Theoretical Physics},\ \bibinfo {editor} {edited by\
  \bibinfo {editor} {\bibfnamefont {J.}~\bibnamefont {{Jedrzejewski}}}}\
  (\bibinfo  {publisher} {{World Scientific}},\ \bibinfo {year} {2015})\ pp.\
  \bibinfo {pages} {1--43},\ \Eprint {http://arxiv.org/abs/1501.00002}
  {arXiv:1501.00002 [cond-mat.str-el]} \BibitemShut {NoStop}%
\bibitem [{\citenamefont {{Sachdev}}\ and\ \citenamefont
  {{Chowdhury}}(2016)}]{SSDC16}%
  \BibitemOpen
  \bibfield  {author} {\bibinfo {author} {\bibfnamefont {S.}~\bibnamefont
  {{Sachdev}}}\ and\ \bibinfo {author} {\bibfnamefont {D.}~\bibnamefont
  {{Chowdhury}}},\ }\bibfield  {title} {\enquote {\bibinfo {title} {{The novel
  metallic states of the cuprates: topological Fermi liquids and strange
  metals}},}\ }\href@noop {} {\bibfield  {journal} {\bibinfo  {journal} {ArXiv
  e-prints}\ } (\bibinfo {year} {2016})},\ \Eprint
  {http://arxiv.org/abs/1605.03579} {arXiv:1605.03579 [cond-mat.str-el]}
  \BibitemShut {NoStop}%
\bibitem [{\citenamefont {{Kaul}}\ \emph {et~al.}(2007)\citenamefont {{Kaul}},
  \citenamefont {{Kolezhuk}}, \citenamefont {{Levin}}, \citenamefont
  {{Sachdev}},\ and\ \citenamefont {{Senthil}}}]{RKK07}%
  \BibitemOpen
  \bibfield  {author} {\bibinfo {author} {\bibfnamefont {R.~K.}\ \bibnamefont
  {{Kaul}}}, \bibinfo {author} {\bibfnamefont {A.}~\bibnamefont {{Kolezhuk}}},
  \bibinfo {author} {\bibfnamefont {M.}~\bibnamefont {{Levin}}}, \bibinfo
  {author} {\bibfnamefont {S.}~\bibnamefont {{Sachdev}}}, \ and\ \bibinfo
  {author} {\bibfnamefont {T.}~\bibnamefont {{Senthil}}},\ }\bibfield  {title}
  {\enquote {\bibinfo {title} {{Hole dynamics in an antiferromagnet across a
  deconfined quantum critical point}},}\ }\href {\doibase
  10.1103/PhysRevB.75.235122} {\bibfield  {journal} {\bibinfo  {journal} {Phys.
  Rev. B}\ }\textbf {\bibinfo {volume} {75}},\ \bibinfo {eid} {235122}
  (\bibinfo {year} {2007})},\ \Eprint {http://arxiv.org/abs/cond-mat/0702119}
  {cond-mat/0702119} \BibitemShut {NoStop}%
\bibitem [{\citenamefont {{Kaul}}\ \emph
  {et~al.}(2008{\natexlab{a}})\citenamefont {{Kaul}}, \citenamefont {{Kim}},
  \citenamefont {{Sachdev}},\ and\ \citenamefont {{Senthil}}}]{RKK08}%
  \BibitemOpen
  \bibfield  {author} {\bibinfo {author} {\bibfnamefont {R.~K.}\ \bibnamefont
  {{Kaul}}}, \bibinfo {author} {\bibfnamefont {Y.~B.}\ \bibnamefont {{Kim}}},
  \bibinfo {author} {\bibfnamefont {S.}~\bibnamefont {{Sachdev}}}, \ and\
  \bibinfo {author} {\bibfnamefont {T.}~\bibnamefont {{Senthil}}},\ }\bibfield
  {title} {\enquote {\bibinfo {title} {{Algebraic charge liquids}},}\ }\href
  {\doibase 10.1038/nphys790} {\bibfield  {journal} {\bibinfo  {journal}
  {Nature Physics}\ }\textbf {\bibinfo {volume} {4}},\ \bibinfo {pages} {28}
  (\bibinfo {year} {2008}{\natexlab{a}})},\ \Eprint
  {http://arxiv.org/abs/0706.2187} {arXiv:0706.2187 [cond-mat.str-el]}
  \BibitemShut {NoStop}%
\bibitem [{\citenamefont {Punk}\ \emph {et~al.}(2015)\citenamefont {Punk},
  \citenamefont {Allais},\ and\ \citenamefont {Sachdev}}]{Punk15}%
  \BibitemOpen
  \bibfield  {author} {\bibinfo {author} {\bibfnamefont {M.}~\bibnamefont
  {Punk}}, \bibinfo {author} {\bibfnamefont {A.}~\bibnamefont {Allais}}, \ and\
  \bibinfo {author} {\bibfnamefont {S.}~\bibnamefont {Sachdev}},\ }\bibfield
  {title} {\enquote {\bibinfo {title} {{A quantum dimer model for the pseudogap
  metal}},}\ }\href {\doibase 10.1073/pnas.1512206112} {\bibfield  {journal}
  {\bibinfo  {journal} {Proc. Nat. Acad. Sci.}\ }\textbf {\bibinfo {volume}
  {112}},\ \bibinfo {pages} {9552} (\bibinfo {year} {2015})},\ \Eprint
  {http://arxiv.org/abs/1501.00978} {arXiv:1501.00978 [cond-mat.str-el]}
  \BibitemShut {NoStop}%
%%CITATION = ARXIV:1501.00978;%%
\bibitem [{\citenamefont {Hertz}(1976)}]{Hertz76}%
  \BibitemOpen
  \bibfield  {author} {\bibinfo {author} {\bibfnamefont {J.~A.}\ \bibnamefont
  {Hertz}},\ }\bibfield  {title} {\enquote {\bibinfo {title} {Quantum critical
  phenomena},}\ }\href {\doibase 10.1103/PhysRevB.14.1165} {\bibfield
  {journal} {\bibinfo  {journal} {Phys. Rev. B}\ }\textbf {\bibinfo {volume}
  {14}},\ \bibinfo {pages} {1165} (\bibinfo {year} {1976})}\BibitemShut
  {NoStop}%
\bibitem [{\citenamefont {Read}\ and\ \citenamefont {Sachdev}(1991)}]{NRSS91}%
  \BibitemOpen
  \bibfield  {author} {\bibinfo {author} {\bibfnamefont {N.}~\bibnamefont
  {Read}}\ and\ \bibinfo {author} {\bibfnamefont {S.}~\bibnamefont {Sachdev}},\
  }\bibfield  {title} {\enquote {\bibinfo {title} {{Large $N$ expansion for
  frustrated quantum antiferromagnets}},}\ }\href {\doibase
  10.1103/PhysRevLett.66.1773} {\bibfield  {journal} {\bibinfo  {journal}
  {Phys. Rev. Lett.}\ }\textbf {\bibinfo {volume} {66}},\ \bibinfo {pages}
  {1773} (\bibinfo {year} {1991})}\BibitemShut {NoStop}%
\bibitem [{\citenamefont {Wen}(1991)}]{XGW91}%
  \BibitemOpen
  \bibfield  {author} {\bibinfo {author} {\bibfnamefont {X.~G.}\ \bibnamefont
  {Wen}},\ }\bibfield  {title} {\enquote {\bibinfo {title} {Mean-field theory
  of spin-liquid states with finite energy gap and topological orders},}\
  }\href {\doibase 10.1103/PhysRevB.44.2664} {\bibfield  {journal} {\bibinfo
  {journal} {Phys. Rev. B}\ }\textbf {\bibinfo {volume} {44}},\ \bibinfo
  {pages} {2664} (\bibinfo {year} {1991})}\BibitemShut {NoStop}%
\bibitem [{\citenamefont {Millis}(1993)}]{Millis93}%
  \BibitemOpen
  \bibfield  {author} {\bibinfo {author} {\bibfnamefont {A.~J.}\ \bibnamefont
  {Millis}},\ }\bibfield  {title} {\enquote {\bibinfo {title} {Effect of a
  nonzero temperature on quantum critical points in itinerant fermion
  systems},}\ }\href {\doibase 10.1103/PhysRevB.48.7183} {\bibfield  {journal}
  {\bibinfo  {journal} {Phys. Rev. B}\ }\textbf {\bibinfo {volume} {48}},\
  \bibinfo {pages} {7183} (\bibinfo {year} {1993})}\BibitemShut {NoStop}%
\bibitem [{\citenamefont {{Abanov}}\ and\ \citenamefont
  {{Chubukov}}(2000)}]{abanov00}%
  \BibitemOpen
  \bibfield  {author} {\bibinfo {author} {\bibfnamefont {A.}~\bibnamefont
  {{Abanov}}}\ and\ \bibinfo {author} {\bibfnamefont {A.~V.}\ \bibnamefont
  {{Chubukov}}},\ }\bibfield  {title} {\enquote {\bibinfo {title}
  {{Spin-Fermion Model near the Quantum Critical Point: One-Loop
  Renormalization Group Results}},}\ }\href {\doibase
  10.1103/PhysRevLett.84.5608} {\bibfield  {journal} {\bibinfo  {journal}
  {Phys. Rev. Lett.}\ }\textbf {\bibinfo {volume} {84}},\ \bibinfo {pages}
  {5608} (\bibinfo {year} {2000})},\ \Eprint
  {http://arxiv.org/abs/cond-mat/0002122} {arXiv:cond-mat/0002122} \BibitemShut
  {NoStop}%
\bibitem [{\citenamefont {Abanov}\ \emph {et~al.}(2003)\citenamefont {Abanov},
  \citenamefont {Chubukov},\ and\ \citenamefont {Schmalian}}]{ACS03}%
  \BibitemOpen
  \bibfield  {author} {\bibinfo {author} {\bibfnamefont {A.}~\bibnamefont
  {Abanov}}, \bibinfo {author} {\bibfnamefont {A.~V.}\ \bibnamefont
  {Chubukov}}, \ and\ \bibinfo {author} {\bibfnamefont {J.}~\bibnamefont
  {Schmalian}},\ }\bibfield  {title} {\enquote {\bibinfo {title}
  {{Quantum-critical theory of the spin-fermion model and its application to
  cuprates: Normal state analysis}},}\ }\href {\doibase
  10.1080/0001873021000057123} {\bibfield  {journal} {\bibinfo  {journal}
  {Advances in Physics}\ }\textbf {\bibinfo {volume} {52}},\ \bibinfo {pages}
  {119} (\bibinfo {year} {2003})},\ \Eprint
  {http://arxiv.org/abs/cond-mat/0107421} {cond-mat/0107421} \BibitemShut
  {NoStop}%
\bibitem [{\citenamefont {{Metlitski}}\ and\ \citenamefont
  {{Sachdev}}(2010)}]{MMSS10b}%
  \BibitemOpen
  \bibfield  {author} {\bibinfo {author} {\bibfnamefont {M.~A.}\ \bibnamefont
  {{Metlitski}}}\ and\ \bibinfo {author} {\bibfnamefont {S.}~\bibnamefont
  {{Sachdev}}},\ }\bibfield  {title} {\enquote {\bibinfo {title} {{Quantum
  phase transitions of metals in two spatial dimensions. II. Spin density wave
  order}},}\ }\href {\doibase 10.1103/PhysRevB.82.075128} {\bibfield  {journal}
  {\bibinfo  {journal} {Phys. Rev. B}\ }\textbf {\bibinfo {volume} {82}},\
  \bibinfo {eid} {075128} (\bibinfo {year} {2010})},\ \Eprint
  {http://arxiv.org/abs/1005.1288} {arXiv:1005.1288 [cond-mat.str-el]}
  \BibitemShut {NoStop}%
\bibitem [{\citenamefont {{Lee}}\ \emph {et~al.}(2013)\citenamefont {{Lee}},
  \citenamefont {{Strack}},\ and\ \citenamefont {{Sachdev}}}]{LeeStrack13}%
  \BibitemOpen
  \bibfield  {author} {\bibinfo {author} {\bibfnamefont {J.}~\bibnamefont
  {{Lee}}}, \bibinfo {author} {\bibfnamefont {P.}~\bibnamefont {{Strack}}}, \
  and\ \bibinfo {author} {\bibfnamefont {S.}~\bibnamefont {{Sachdev}}},\
  }\bibfield  {title} {\enquote {\bibinfo {title} {{Quantum criticality of
  reconstructing Fermi surfaces in antiferromagnetic metals}},}\ }\href
  {\doibase 10.1103/PhysRevB.87.045104} {\bibfield  {journal} {\bibinfo
  {journal} {Phys. Rev. B}\ }\textbf {\bibinfo {volume} {87}},\ \bibinfo {eid}
  {045104} (\bibinfo {year} {2013})},\ \Eprint {http://arxiv.org/abs/1209.4644}
  {arXiv:1209.4644 [cond-mat.str-el]} \BibitemShut {NoStop}%
\bibitem [{\citenamefont {{Sur}}\ and\ \citenamefont {{Lee}}(2015)}]{SurLee15}%
  \BibitemOpen
  \bibfield  {author} {\bibinfo {author} {\bibfnamefont {S.}~\bibnamefont
  {{Sur}}}\ and\ \bibinfo {author} {\bibfnamefont {S.-S.}\ \bibnamefont
  {{Lee}}},\ }\bibfield  {title} {\enquote {\bibinfo {title} {{Quasilocal
  strange metal}},}\ }\href {\doibase 10.1103/PhysRevB.91.125136} {\bibfield
  {journal} {\bibinfo  {journal} {Phys. Rev. B}\ }\textbf {\bibinfo {volume}
  {91}},\ \bibinfo {eid} {125136} (\bibinfo {year} {2015})},\ \Eprint
  {http://arxiv.org/abs/1405.7357} {arXiv:1405.7357 [cond-mat.str-el]}
  \BibitemShut {NoStop}%
\bibitem [{\citenamefont {{Patel}}\ \emph {et~al.}(2015)\citenamefont
  {{Patel}}, \citenamefont {{Strack}},\ and\ \citenamefont
  {{Sachdev}}}]{PatelStrack15}%
  \BibitemOpen
  \bibfield  {author} {\bibinfo {author} {\bibfnamefont {A.~A.}\ \bibnamefont
  {{Patel}}}, \bibinfo {author} {\bibfnamefont {P.}~\bibnamefont {{Strack}}}, \
  and\ \bibinfo {author} {\bibfnamefont {S.}~\bibnamefont {{Sachdev}}},\
  }\bibfield  {title} {\enquote {\bibinfo {title} {{Hyperscaling at the spin
  density wave quantum critical point in two-dimensional metals}},}\ }\href
  {\doibase 10.1103/PhysRevB.92.165105} {\bibfield  {journal} {\bibinfo
  {journal} {Phys. Rev. B}\ }\textbf {\bibinfo {volume} {92}},\ \bibinfo {eid}
  {165105} (\bibinfo {year} {2015})},\ \Eprint
  {http://arxiv.org/abs/1507.05962} {arXiv:1507.05962 [cond-mat.str-el]}
  \BibitemShut {NoStop}%
\bibitem [{\citenamefont {{Maier}}\ and\ \citenamefont
  {{Strack}}(2016)}]{Strack16}%
  \BibitemOpen
  \bibfield  {author} {\bibinfo {author} {\bibfnamefont {S.~A.}\ \bibnamefont
  {{Maier}}}\ and\ \bibinfo {author} {\bibfnamefont {P.}~\bibnamefont
  {{Strack}}},\ }\bibfield  {title} {\enquote {\bibinfo {title} {{Universality
  in antiferromagnetic strange metals}},}\ }\href {\doibase
  10.1103/PhysRevB.93.165114} {\bibfield  {journal} {\bibinfo  {journal} {Phys.
  Rev. B}\ }\textbf {\bibinfo {volume} {93}},\ \bibinfo {eid} {165114}
  (\bibinfo {year} {2016})},\ \Eprint {http://arxiv.org/abs/1510.01331}
  {arXiv:1510.01331 [cond-mat.str-el]} \BibitemShut {NoStop}%
\bibitem [{\citenamefont {{Berg}}\ \emph {et~al.}(2012)\citenamefont {{Berg}},
  \citenamefont {{Metlitski}},\ and\ \citenamefont {{Sachdev}}}]{BMS12}%
  \BibitemOpen
  \bibfield  {author} {\bibinfo {author} {\bibfnamefont {E.}~\bibnamefont
  {{Berg}}}, \bibinfo {author} {\bibfnamefont {M.~A.}\ \bibnamefont
  {{Metlitski}}}, \ and\ \bibinfo {author} {\bibfnamefont {S.}~\bibnamefont
  {{Sachdev}}},\ }\bibfield  {title} {\enquote {\bibinfo {title}
  {{Sign-Problem-Free Quantum Monte Carlo of the Onset of Antiferromagnetism in
  Metals}},}\ }\href {\doibase 10.1126/science.1227769} {\bibfield  {journal}
  {\bibinfo  {journal} {Science}\ }\textbf {\bibinfo {volume} {338}},\ \bibinfo
  {pages} {1606} (\bibinfo {year} {2012})},\ \Eprint
  {http://arxiv.org/abs/1206.0742} {arXiv:1206.0742 [cond-mat.str-el]}
  \BibitemShut {NoStop}%
\bibitem [{\citenamefont {Schattner}\ \emph {et~al.}(2016)\citenamefont
  {Schattner}, \citenamefont {Gerlach}, \citenamefont {Trebst},\ and\
  \citenamefont {Berg}}]{SGTB15}%
  \BibitemOpen
  \bibfield  {author} {\bibinfo {author} {\bibfnamefont {Y.}~\bibnamefont
  {Schattner}}, \bibinfo {author} {\bibfnamefont {M.~H.}\ \bibnamefont
  {Gerlach}}, \bibinfo {author} {\bibfnamefont {S.}~\bibnamefont {Trebst}}, \
  and\ \bibinfo {author} {\bibfnamefont {E.}~\bibnamefont {Berg}},\ }\bibfield
  {title} {\enquote {\bibinfo {title} {Competing orders in a nearly
  antiferromagnetic metal},}\ }\href {\doibase 10.1103/PhysRevLett.117.097002}
  {\bibfield  {journal} {\bibinfo  {journal} {Phys. Rev. Lett.}\ }\textbf
  {\bibinfo {volume} {117}},\ \bibinfo {pages} {097002} (\bibinfo {year}
  {2016})},\ \Eprint {http://arxiv.org/abs/1512.07257} {arXiv:1512.07257
  [cond-mat.supr-con]} \BibitemShut {NoStop}%
\bibitem [{\citenamefont {{Li}}\ \emph {et~al.}(2015)\citenamefont {{Li}},
  \citenamefont {{Wang}}, \citenamefont {{Yao}},\ and\ \citenamefont
  {{Lee}}}]{LWYL15}%
  \BibitemOpen
  \bibfield  {author} {\bibinfo {author} {\bibfnamefont {Z.-X.}\ \bibnamefont
  {{Li}}}, \bibinfo {author} {\bibfnamefont {F.}~\bibnamefont {{Wang}}},
  \bibinfo {author} {\bibfnamefont {H.}~\bibnamefont {{Yao}}}, \ and\ \bibinfo
  {author} {\bibfnamefont {D.-H.}\ \bibnamefont {{Lee}}},\ }\bibfield  {title}
  {\enquote {\bibinfo {title} {{The nature of effective interaction in cuprate
  superconductors: a sign-problem-free quantum Monte-Carlo study}},}\
  }\href@noop {} {\bibfield  {journal} {\bibinfo  {journal} {ArXiv e-prints}\ }
  (\bibinfo {year} {2015})},\ \Eprint {http://arxiv.org/abs/1512.04541}
  {arXiv:1512.04541 [cond-mat.supr-con]} \BibitemShut {NoStop}%
\bibitem [{\citenamefont {Shraiman}\ and\ \citenamefont {Siggia}(1988)}]{SS88}%
  \BibitemOpen
  \bibfield  {author} {\bibinfo {author} {\bibfnamefont {B.~I.}\ \bibnamefont
  {Shraiman}}\ and\ \bibinfo {author} {\bibfnamefont {E.~D.}\ \bibnamefont
  {Siggia}},\ }\bibfield  {title} {\enquote {\bibinfo {title} {{Mobile
  Vacancies in a Quantum Heisenberg Antiferromagnet}},}\ }\href {\doibase
  10.1103/PhysRevLett.61.467} {\bibfield  {journal} {\bibinfo  {journal} {Phys.
  Rev. Lett.}\ }\textbf {\bibinfo {volume} {61}},\ \bibinfo {pages} {467}
  (\bibinfo {year} {1988})}\BibitemShut {NoStop}%
\bibitem [{\citenamefont {Shraiman}\ and\ \citenamefont {Siggia}(1989)}]{SS89}%
  \BibitemOpen
  \bibfield  {author} {\bibinfo {author} {\bibfnamefont {B.~I.}\ \bibnamefont
  {Shraiman}}\ and\ \bibinfo {author} {\bibfnamefont {E.~D.}\ \bibnamefont
  {Siggia}},\ }\bibfield  {title} {\enquote {\bibinfo {title} {{Spiral phase of
  a doped quantum antiferromagnet}},}\ }\href {\doibase
  10.1103/PhysRevLett.62.1564} {\bibfield  {journal} {\bibinfo  {journal}
  {Phys. Rev. Lett.}\ }\textbf {\bibinfo {volume} {62}},\ \bibinfo {pages}
  {1564} (\bibinfo {year} {1989})}\BibitemShut {NoStop}%
\bibitem [{\citenamefont {{Senthil}}\ and\ \citenamefont
  {{Fisher}}(2000)}]{TSMPAF00}%
  \BibitemOpen
  \bibfield  {author} {\bibinfo {author} {\bibfnamefont {T.}~\bibnamefont
  {{Senthil}}}\ and\ \bibinfo {author} {\bibfnamefont {M.~P.~A.}\ \bibnamefont
  {{Fisher}}},\ }\bibfield  {title} {\enquote {\bibinfo {title}
  {{$\mathbb{Z}_{2}$ gauge theory of electron fractionalization in strongly
  correlated systems}},}\ }\href {\doibase 10.1103/PhysRevB.62.7850} {\bibfield
   {journal} {\bibinfo  {journal} {Phys. Rev. B}\ }\textbf {\bibinfo {volume}
  {62}},\ \bibinfo {pages} {7850} (\bibinfo {year} {2000})},\ \Eprint
  {http://arxiv.org/abs/cond-mat/9910224} {cond-mat/9910224} \BibitemShut
  {NoStop}%
\bibitem [{\citenamefont {{Sedgewick}}\ \emph {et~al.}(2002)\citenamefont
  {{Sedgewick}}, \citenamefont {{Scalapino}},\ and\ \citenamefont
  {{Sugar}}}]{SSS02}%
  \BibitemOpen
  \bibfield  {author} {\bibinfo {author} {\bibfnamefont {R.~D.}\ \bibnamefont
  {{Sedgewick}}}, \bibinfo {author} {\bibfnamefont {D.~J.}\ \bibnamefont
  {{Scalapino}}}, \ and\ \bibinfo {author} {\bibfnamefont {R.~L.}\ \bibnamefont
  {{Sugar}}},\ }\bibfield  {title} {\enquote {\bibinfo {title} {{Fractionalized
  phase in an XY-$\mathbb{Z}_2$ gauge model}},}\ }\href {\doibase
  10.1103/PhysRevB.65.054508} {\bibfield  {journal} {\bibinfo  {journal} {Phys.
  Rev. B}\ }\textbf {\bibinfo {volume} {65}},\ \bibinfo {eid} {054508}
  (\bibinfo {year} {2002})},\ \Eprint {http://arxiv.org/abs/cond-mat/0012028}
  {cond-mat/0012028} \BibitemShut {NoStop}%
\bibitem [{\citenamefont {{Sachdev}}\ and\ \citenamefont
  {{Park}}(2002)}]{KPSS02}%
  \BibitemOpen
  \bibfield  {author} {\bibinfo {author} {\bibfnamefont {S.}~\bibnamefont
  {{Sachdev}}}\ and\ \bibinfo {author} {\bibfnamefont {K.}~\bibnamefont
  {{Park}}},\ }\bibfield  {title} {\enquote {\bibinfo {title} {{Ground States
  of Quantum Antiferromagnets in Two Dimensions}},}\ }\href {\doibase
  10.1006/aphy.2002.6232} {\bibfield  {journal} {\bibinfo  {journal} {Annals of
  Physics}\ }\textbf {\bibinfo {volume} {298}},\ \bibinfo {pages} {58}
  (\bibinfo {year} {2002})},\ \Eprint {http://arxiv.org/abs/cond-mat/0108214}
  {cond-mat/0108214} \BibitemShut {NoStop}%
\bibitem [{\citenamefont {{Park}}\ and\ \citenamefont
  {{Sachdev}}(2002)}]{KPSS02a}%
  \BibitemOpen
  \bibfield  {author} {\bibinfo {author} {\bibfnamefont {K.}~\bibnamefont
  {{Park}}}\ and\ \bibinfo {author} {\bibfnamefont {S.}~\bibnamefont
  {{Sachdev}}},\ }\bibfield  {title} {\enquote {\bibinfo {title} {{Bond and
  N{\'e}el order and fractionalization in ground states of easy-plane
  antiferromagnets in two dimensions}},}\ }\href {\doibase
  10.1103/PhysRevB.65.220405} {\bibfield  {journal} {\bibinfo  {journal} {Phys.
  Rev. B}\ }\textbf {\bibinfo {volume} {65}},\ \bibinfo {eid} {220405}
  (\bibinfo {year} {2002})},\ \Eprint {http://arxiv.org/abs/cond-mat/0112003}
  {cond-mat/0112003} \BibitemShut {NoStop}%
\bibitem [{\citenamefont {Sachdev}\ and\ \citenamefont {Read}(1991)}]{SSNR91}%
  \BibitemOpen
  \bibfield  {author} {\bibinfo {author} {\bibfnamefont {S.}~\bibnamefont
  {Sachdev}}\ and\ \bibinfo {author} {\bibfnamefont {N.}~\bibnamefont {Read}},\
  }\bibfield  {title} {\enquote {\bibinfo {title} {{Large $N$ expansion for
  frustrated and doped quantum antiferromagnets}},}\ }\href {\doibase
  10.1142/S0217979291000158} {\bibfield  {journal} {\bibinfo  {journal} {Int.
  J. Mod. Phys. B}\ }\textbf {\bibinfo {volume} {5}},\ \bibinfo {pages} {219}
  (\bibinfo {year} {1991})},\ \Eprint {http://arxiv.org/abs/cond-mat/0402109}
  {cond-mat/0402109} \BibitemShut {NoStop}%
\bibitem [{\citenamefont {{Chubukov}}\ \emph {et~al.}(1994)\citenamefont
  {{Chubukov}}, \citenamefont {{Senthil}},\ and\ \citenamefont
  {{Sachdev}}}]{CSS94}%
  \BibitemOpen
  \bibfield  {author} {\bibinfo {author} {\bibfnamefont {A.~V.}\ \bibnamefont
  {{Chubukov}}}, \bibinfo {author} {\bibfnamefont {T.}~\bibnamefont
  {{Senthil}}}, \ and\ \bibinfo {author} {\bibfnamefont {S.}~\bibnamefont
  {{Sachdev}}},\ }\bibfield  {title} {\enquote {\bibinfo {title} {{Universal
  magnetic properties of frustrated quantum antiferromagnets in two
  dimensions}},}\ }\href {\doibase 10.1103/PhysRevLett.72.2089} {\bibfield
  {journal} {\bibinfo  {journal} {Phys. Rev. Lett.}\ }\textbf {\bibinfo
  {volume} {72}},\ \bibinfo {pages} {2089} (\bibinfo {year} {1994})},\ \Eprint
  {http://arxiv.org/abs/cond-mat/9311045} {cond-mat/9311045} \BibitemShut
  {NoStop}%
\bibitem [{\citenamefont {Lee}\ \emph {et~al.}(2006)\citenamefont {Lee},
  \citenamefont {Nagaosa},\ and\ \citenamefont {Wen}}]{LeeWenRMP}%
  \BibitemOpen
  \bibfield  {author} {\bibinfo {author} {\bibfnamefont {P.~A.}\ \bibnamefont
  {Lee}}, \bibinfo {author} {\bibfnamefont {N.}~\bibnamefont {Nagaosa}}, \ and\
  \bibinfo {author} {\bibfnamefont {X.-G.}\ \bibnamefont {Wen}},\ }\bibfield
  {title} {\enquote {\bibinfo {title} {{Doping a Mott insulator: Physics of
  high-temperature superconductivity}},}\ }\href {\doibase
  10.1103/RevModPhys.78.17} {\bibfield  {journal} {\bibinfo  {journal} {Rev.
  Mod. Phys.}\ }\textbf {\bibinfo {volume} {78}},\ \bibinfo {pages} {17}
  (\bibinfo {year} {2006})},\ \Eprint {http://arxiv.org/abs/cond-mat/0410445}
  {cond-mat/0410445} \BibitemShut {NoStop}%
\bibitem [{\citenamefont {{Florens}}\ and\ \citenamefont
  {{Georges}}(2004)}]{FG04}%
  \BibitemOpen
  \bibfield  {author} {\bibinfo {author} {\bibfnamefont {S.}~\bibnamefont
  {{Florens}}}\ and\ \bibinfo {author} {\bibfnamefont {A.}~\bibnamefont
  {{Georges}}},\ }\bibfield  {title} {\enquote {\bibinfo {title} {{Slave-rotor
  mean-field theories of strongly correlated systems and the Mott transitionin
  finite dimensions}},}\ }\href {\doibase 10.1103/PhysRevB.70.035114}
  {\bibfield  {journal} {\bibinfo  {journal} {Phys. Rev. B}\ }\textbf {\bibinfo
  {volume} {70}},\ \bibinfo {eid} {035114} (\bibinfo {year} {2004})},\ \Eprint
  {http://arxiv.org/abs/cond-mat/0404334} {cond-mat/0404334} \BibitemShut
  {NoStop}%
\bibitem [{\citenamefont {Lee}(1989)}]{PAL89}%
  \BibitemOpen
  \bibfield  {author} {\bibinfo {author} {\bibfnamefont {P.~A.}\ \bibnamefont
  {Lee}},\ }\bibfield  {title} {\enquote {\bibinfo {title} {{Gauge field,
  Aharonov-Bohm flux, and high-${T}_{c}$ superconductivity}},}\ }\href
  {\doibase 10.1103/PhysRevLett.63.680} {\bibfield  {journal} {\bibinfo
  {journal} {Phys. Rev. Lett.}\ }\textbf {\bibinfo {volume} {63}},\ \bibinfo
  {pages} {680} (\bibinfo {year} {1989})}\BibitemShut {NoStop}%
\bibitem [{\citenamefont {{Qi}}\ and\ \citenamefont
  {{Sachdev}}(2010)}]{YQSS10}%
  \BibitemOpen
  \bibfield  {author} {\bibinfo {author} {\bibfnamefont {Y.}~\bibnamefont
  {{Qi}}}\ and\ \bibinfo {author} {\bibfnamefont {S.}~\bibnamefont
  {{Sachdev}}},\ }\bibfield  {title} {\enquote {\bibinfo {title} {{Effective
  theory of Fermi pockets in fluctuating antiferromagnets}},}\ }\href {\doibase
  10.1103/PhysRevB.81.115129} {\bibfield  {journal} {\bibinfo  {journal} {Phys.
  Rev. B}\ }\textbf {\bibinfo {volume} {81}},\ \bibinfo {eid} {115129}
  (\bibinfo {year} {2010})},\ \Eprint {http://arxiv.org/abs/0912.0943}
  {arXiv:0912.0943 [cond-mat.str-el]} \BibitemShut {NoStop}%
\bibitem [{\citenamefont {{Mei}}\ \emph {et~al.}(2012)\citenamefont {{Mei}},
  \citenamefont {{Kawasaki}}, \citenamefont {{Zheng}}, \citenamefont {{Weng}},\
  and\ \citenamefont {{Wen}}}]{Mei12}%
  \BibitemOpen
  \bibfield  {author} {\bibinfo {author} {\bibfnamefont {J.-W.}\ \bibnamefont
  {{Mei}}}, \bibinfo {author} {\bibfnamefont {S.}~\bibnamefont {{Kawasaki}}},
  \bibinfo {author} {\bibfnamefont {G.-Q.}\ \bibnamefont {{Zheng}}}, \bibinfo
  {author} {\bibfnamefont {Z.-Y.}\ \bibnamefont {{Weng}}}, \ and\ \bibinfo
  {author} {\bibfnamefont {X.-G.}\ \bibnamefont {{Wen}}},\ }\bibfield  {title}
  {\enquote {\bibinfo {title} {{Luttinger-volume violating Fermi liquid in the
  pseudogap phase of the cuprate superconductors}},}\ }\href {\doibase
  10.1103/PhysRevB.85.134519} {\bibfield  {journal} {\bibinfo  {journal} {Phys.
  Rev. B}\ }\textbf {\bibinfo {volume} {85}},\ \bibinfo {eid} {134519}
  (\bibinfo {year} {2012})},\ \Eprint {http://arxiv.org/abs/1109.0406}
  {arXiv:1109.0406 [cond-mat.supr-con]} \BibitemShut {NoStop}%
\bibitem [{\citenamefont {{Tsvelik}}(2016)}]{AT16}%
  \BibitemOpen
  \bibfield  {author} {\bibinfo {author} {\bibfnamefont {A.~M.}\ \bibnamefont
  {{Tsvelik}}},\ }\bibfield  {title} {\enquote {\bibinfo {title}
  {{Fractionalized Fermi Liquid in a Kondo-Heisenberg Model}},}\ }\href@noop {}
  {\bibfield  {journal} {\bibinfo  {journal} {ArXiv e-prints}\ } (\bibinfo
  {year} {2016})},\ \Eprint {http://arxiv.org/abs/1604.06417} {arXiv:1604.06417
  [cond-mat.str-el]} \BibitemShut {NoStop}%
\bibitem [{\citenamefont {{Goldstein}}\ \emph {et~al.}(2016)\citenamefont
  {{Goldstein}}, \citenamefont {{Chamon}},\ and\ \citenamefont
  {{Castelnovo}}}]{GCC16}%
  \BibitemOpen
  \bibfield  {author} {\bibinfo {author} {\bibfnamefont {G.}~\bibnamefont
  {{Goldstein}}}, \bibinfo {author} {\bibfnamefont {C.}~\bibnamefont
  {{Chamon}}}, \ and\ \bibinfo {author} {\bibfnamefont {C.}~\bibnamefont
  {{Castelnovo}}},\ }\bibfield  {title} {\enquote {\bibinfo {title} {{D-wave
  superconductivity in boson+fermion dimer models}},}\ }\href@noop {}
  {\bibfield  {journal} {\bibinfo  {journal} {ArXiv e-prints}\ } (\bibinfo
  {year} {2016})},\ \Eprint {http://arxiv.org/abs/1606.04129} {arXiv:1606.04129
  [cond-mat.str-el]} \BibitemShut {NoStop}%
\bibitem [{\citenamefont {{Lee}}\ \emph {et~al.}(2016)\citenamefont {{Lee}},
  \citenamefont {{Sachdev}},\ and\ \citenamefont {{White}}}]{LSW16}%
  \BibitemOpen
  \bibfield  {author} {\bibinfo {author} {\bibfnamefont {J.}~\bibnamefont
  {{Lee}}}, \bibinfo {author} {\bibfnamefont {S.}~\bibnamefont {{Sachdev}}}, \
  and\ \bibinfo {author} {\bibfnamefont {S.~R.}\ \bibnamefont {{White}}},\
  }\bibfield  {title} {\enquote {\bibinfo {title} {{Electronic quasiparticles
  in the quantum dimer model: density matrix renormalization group results}},}\
  }\href@noop {} {\bibfield  {journal} {\bibinfo  {journal} {ArXiv e-prints}\ }
  (\bibinfo {year} {2016})},\ \Eprint {http://arxiv.org/abs/1606.04105}
  {arXiv:1606.04105 [cond-mat.str-el]} \BibitemShut {NoStop}%
\bibitem [{\citenamefont {{Grover}}\ and\ \citenamefont
  {{Senthil}}(2010)}]{TGTS10}%
  \BibitemOpen
  \bibfield  {author} {\bibinfo {author} {\bibfnamefont {T.}~\bibnamefont
  {{Grover}}}\ and\ \bibinfo {author} {\bibfnamefont {T.}~\bibnamefont
  {{Senthil}}},\ }\bibfield  {title} {\enquote {\bibinfo {title} {{Quantum
  phase transition from an antiferromagnet to a spin liquid in a metal}},}\
  }\href {\doibase 10.1103/PhysRevB.81.205102} {\bibfield  {journal} {\bibinfo
  {journal} {Phys. Rev. B}\ }\textbf {\bibinfo {volume} {81}},\ \bibinfo {eid}
  {205102} (\bibinfo {year} {2010})},\ \Eprint {http://arxiv.org/abs/0910.1277}
  {arXiv:0910.1277 [cond-mat.str-el]} \BibitemShut {NoStop}%
\bibitem [{\citenamefont {Jalabert}\ and\ \citenamefont
  {Sachdev}(1991)}]{RJSS91}%
  \BibitemOpen
  \bibfield  {author} {\bibinfo {author} {\bibfnamefont {R.~A.}\ \bibnamefont
  {Jalabert}}\ and\ \bibinfo {author} {\bibfnamefont {S.}~\bibnamefont
  {Sachdev}},\ }\bibfield  {title} {\enquote {\bibinfo {title} {{Spontaneous
  alignment of frustrated bonds in an anisotropic, three-dimensional Ising
  model}},}\ }\href {\doibase 10.1103/PhysRevB.44.686} {\bibfield  {journal}
  {\bibinfo  {journal} {Phys. Rev. B}\ }\textbf {\bibinfo {volume} {44}},\
  \bibinfo {pages} {686} (\bibinfo {year} {1991})}\BibitemShut {NoStop}%
\bibitem [{\citenamefont {{Sachdev}}\ and\ \citenamefont
  {{Vojta}}(1999)}]{MVSS99}%
  \BibitemOpen
  \bibfield  {author} {\bibinfo {author} {\bibfnamefont {S.}~\bibnamefont
  {{Sachdev}}}\ and\ \bibinfo {author} {\bibfnamefont {M.}~\bibnamefont
  {{Vojta}}},\ }\bibfield  {title} {\enquote {\bibinfo {title} {{Translational
  symmetry breaking in two-dimensional antiferromagnets and
  superconductors}},}\ }\href@noop {} {\bibfield  {journal} {\bibinfo
  {journal} {J. Phys. Soc. Jpn {\bf 69}, Supp. B, 1}\ } (\bibinfo {year}
  {1999})},\ \Eprint {http://arxiv.org/abs/cond-mat/9910231} {cond-mat/9910231}
  \BibitemShut {NoStop}%
\bibitem [{\citenamefont {{Moessner}}\ \emph {et~al.}(2001)\citenamefont
  {{Moessner}}, \citenamefont {{Sondhi}},\ and\ \citenamefont
  {{Fradkin}}}]{MSF02}%
  \BibitemOpen
  \bibfield  {author} {\bibinfo {author} {\bibfnamefont {R.}~\bibnamefont
  {{Moessner}}}, \bibinfo {author} {\bibfnamefont {S.~L.}\ \bibnamefont
  {{Sondhi}}}, \ and\ \bibinfo {author} {\bibfnamefont {E.}~\bibnamefont
  {{Fradkin}}},\ }\bibfield  {title} {\enquote {\bibinfo {title} {{Short-ranged
  resonating valence bond physics, quantum dimer models, and Ising gauge
  theories}},}\ }\href {\doibase 10.1103/PhysRevB.65.024504} {\bibfield
  {journal} {\bibinfo  {journal} {Phys. Rev. B}\ }\textbf {\bibinfo {volume}
  {65}},\ \bibinfo {eid} {024504} (\bibinfo {year} {2001})},\ \Eprint
  {http://arxiv.org/abs/cond-mat/0103396} {cond-mat/0103396} \BibitemShut
  {NoStop}%
\bibitem [{\citenamefont {{Kitaev}}(2003)}]{Kitaev03}%
  \BibitemOpen
  \bibfield  {author} {\bibinfo {author} {\bibfnamefont {A.~Y.}\ \bibnamefont
  {{Kitaev}}},\ }\bibfield  {title} {\enquote {\bibinfo {title}
  {{Fault-tolerant quantum computation by anyons}},}\ }\href {\doibase
  10.1016/S0003-4916(02)00018-0} {\bibfield  {journal} {\bibinfo  {journal}
  {Annals of Physics}\ }\textbf {\bibinfo {volume} {303}},\ \bibinfo {pages}
  {2} (\bibinfo {year} {2003})},\ \Eprint
  {http://arxiv.org/abs/quant-ph/9707021} {quant-ph/9707021} \BibitemShut
  {NoStop}%
\bibitem [{\citenamefont {Kogut}(1979)}]{KogutRMP}%
  \BibitemOpen
  \bibfield  {author} {\bibinfo {author} {\bibfnamefont {J.~B.}\ \bibnamefont
  {Kogut}},\ }\bibfield  {title} {\enquote {\bibinfo {title} {An introduction
  to lattice gauge theory and spin systems},}\ }\href {\doibase
  10.1103/RevModPhys.51.659} {\bibfield  {journal} {\bibinfo  {journal} {Rev.
  Mod. Phys.}\ }\textbf {\bibinfo {volume} {51}},\ \bibinfo {pages} {659}
  (\bibinfo {year} {1979})}\BibitemShut {NoStop}%
\bibitem [{\citenamefont {{Huh}}\ \emph {et~al.}(2011)\citenamefont {{Huh}},
  \citenamefont {{Punk}},\ and\ \citenamefont {{Sachdev}}}]{YHMPSS11}%
  \BibitemOpen
  \bibfield  {author} {\bibinfo {author} {\bibfnamefont {Y.}~\bibnamefont
  {{Huh}}}, \bibinfo {author} {\bibfnamefont {M.}~\bibnamefont {{Punk}}}, \
  and\ \bibinfo {author} {\bibfnamefont {S.}~\bibnamefont {{Sachdev}}},\
  }\bibfield  {title} {\enquote {\bibinfo {title} {{Vison states and
  confinement transitions of $\mathbb{Z}_{2}$ spin liquids on the kagome
  lattice}},}\ }\href {\doibase 10.1103/PhysRevB.84.094419} {\bibfield
  {journal} {\bibinfo  {journal} {Phys. Rev. B}\ }\textbf {\bibinfo {volume}
  {84}},\ \bibinfo {eid} {094419} (\bibinfo {year} {2011})},\ \Eprint
  {http://arxiv.org/abs/1106.3330} {arXiv:1106.3330 [cond-mat.str-el]}
  \BibitemShut {NoStop}%
\bibitem [{\citenamefont {{Patel}}\ \emph {et~al.}(2016)\citenamefont
  {{Patel}}, \citenamefont {{Chowdhury}}, \citenamefont {{Allais}},\ and\
  \citenamefont {{Sachdev}}}]{PCAS16}%
  \BibitemOpen
  \bibfield  {author} {\bibinfo {author} {\bibfnamefont {A.~A.}\ \bibnamefont
  {{Patel}}}, \bibinfo {author} {\bibfnamefont {D.}~\bibnamefont
  {{Chowdhury}}}, \bibinfo {author} {\bibfnamefont {A.}~\bibnamefont
  {{Allais}}}, \ and\ \bibinfo {author} {\bibfnamefont {S.}~\bibnamefont
  {{Sachdev}}},\ }\bibfield  {title} {\enquote {\bibinfo {title} {{Confinement
  transition to density wave order in metallic doped spin liquids}},}\ }\href
  {\doibase 10.1103/PhysRevB.93.165139} {\bibfield  {journal} {\bibinfo
  {journal} {Phys. Rev. B}\ }\textbf {\bibinfo {volume} {93}},\ \bibinfo {eid}
  {165139} (\bibinfo {year} {2016})},\ \Eprint
  {http://arxiv.org/abs/1602.05954} {arXiv:1602.05954 [cond-mat.str-el]}
  \BibitemShut {NoStop}%
\bibitem [{\citenamefont {{Zhang}}\ \emph {et~al.}(2002)\citenamefont
  {{Zhang}}, \citenamefont {{Demler}},\ and\ \citenamefont
  {{Sachdev}}}]{ZDS02}%
  \BibitemOpen
  \bibfield  {author} {\bibinfo {author} {\bibfnamefont {Y.}~\bibnamefont
  {{Zhang}}}, \bibinfo {author} {\bibfnamefont {E.}~\bibnamefont {{Demler}}}, \
  and\ \bibinfo {author} {\bibfnamefont {S.}~\bibnamefont {{Sachdev}}},\
  }\bibfield  {title} {\enquote {\bibinfo {title} {{Competing orders in a
  magnetic field: Spin and charge order in the cuprate superconductors}},}\
  }\href {\doibase 10.1103/PhysRevB.66.094501} {\bibfield  {journal} {\bibinfo
  {journal} {Phys. Rev. B}\ }\textbf {\bibinfo {volume} {66}},\ \bibinfo {eid}
  {094501} (\bibinfo {year} {2002})},\ \Eprint
  {http://arxiv.org/abs/cond-mat/0112343} {cond-mat/0112343} \BibitemShut
  {NoStop}%
\bibitem [{\citenamefont {{Sachdev}}(2003)}]{SSRMP}%
  \BibitemOpen
  \bibfield  {author} {\bibinfo {author} {\bibfnamefont {S.}~\bibnamefont
  {{Sachdev}}},\ }\bibfield  {title} {\enquote {\bibinfo {title} {{Colloquium:
  Order and quantum phase transitions in the cuprate superconductors}},}\
  }\href {\doibase 10.1103/RevModPhys.75.913} {\bibfield  {journal} {\bibinfo
  {journal} {Rev. Mod. Phys.}\ }\textbf {\bibinfo {volume} {75}},\ \bibinfo
  {pages} {913} (\bibinfo {year} {2003})},\ \Eprint
  {http://arxiv.org/abs/cond-mat/0211005} {cond-mat/0211005} \BibitemShut
  {NoStop}%
\bibitem [{\citenamefont {Read}\ and\ \citenamefont {Sachdev}(1989)}]{NRSS89}%
  \BibitemOpen
  \bibfield  {author} {\bibinfo {author} {\bibfnamefont {N.}~\bibnamefont
  {Read}}\ and\ \bibinfo {author} {\bibfnamefont {S.}~\bibnamefont {Sachdev}},\
  }\bibfield  {title} {\enquote {\bibinfo {title} {{Valence-bond and
  spin-Peierls ground states of low-dimensional quantum antiferromagnets}},}\
  }\href {\doibase 10.1103/PhysRevLett.62.1694} {\bibfield  {journal} {\bibinfo
   {journal} {Phys. Rev. Lett.}\ }\textbf {\bibinfo {volume} {62}},\ \bibinfo
  {pages} {1694} (\bibinfo {year} {1989})}\BibitemShut {NoStop}%
\bibitem [{\citenamefont {Read}\ and\ \citenamefont {Sachdev}(1990)}]{NRSS90}%
  \BibitemOpen
  \bibfield  {author} {\bibinfo {author} {\bibfnamefont {N.}~\bibnamefont
  {Read}}\ and\ \bibinfo {author} {\bibfnamefont {S.}~\bibnamefont {Sachdev}},\
  }\bibfield  {title} {\enquote {\bibinfo {title} {{Spin-Peierls, valence-bond
  solid, and N\'eel ground states of low-dimensional quantum
  antiferromagnets}},}\ }\href {\doibase 10.1103/PhysRevB.42.4568} {\bibfield
  {journal} {\bibinfo  {journal} {Phys. Rev. B}\ }\textbf {\bibinfo {volume}
  {42}},\ \bibinfo {pages} {4568} (\bibinfo {year} {1990})}\BibitemShut
  {NoStop}%
\bibitem [{\citenamefont {{Fu}}\ \emph {et~al.}(2011)\citenamefont {{Fu}},
  \citenamefont {{Sachdev}},\ and\ \citenamefont {{Xu}}}]{FSX11}%
  \BibitemOpen
  \bibfield  {author} {\bibinfo {author} {\bibfnamefont {L.}~\bibnamefont
  {{Fu}}}, \bibinfo {author} {\bibfnamefont {S.}~\bibnamefont {{Sachdev}}}, \
  and\ \bibinfo {author} {\bibfnamefont {C.}~\bibnamefont {{Xu}}},\ }\bibfield
  {title} {\enquote {\bibinfo {title} {{Geometric phases and competing orders
  in two dimensions}},}\ }\href {\doibase 10.1103/PhysRevB.83.165123}
  {\bibfield  {journal} {\bibinfo  {journal} {Phys. Rev. B}\ }\textbf {\bibinfo
  {volume} {83}},\ \bibinfo {eid} {165123} (\bibinfo {year} {2011})},\ \Eprint
  {http://arxiv.org/abs/1010.3745} {arXiv:1010.3745 [cond-mat.str-el]}
  \BibitemShut {NoStop}%
\bibitem [{\citenamefont {{Hermele}}\ \emph {et~al.}(2004)\citenamefont
  {{Hermele}}, \citenamefont {{Senthil}}, \citenamefont {{Fisher}},
  \citenamefont {{Lee}}, \citenamefont {{Nagaosa}},\ and\ \citenamefont
  {{Wen}}}]{Hermele04}%
  \BibitemOpen
  \bibfield  {author} {\bibinfo {author} {\bibfnamefont {M.}~\bibnamefont
  {{Hermele}}}, \bibinfo {author} {\bibfnamefont {T.}~\bibnamefont
  {{Senthil}}}, \bibinfo {author} {\bibfnamefont {M.~P.~A.}\ \bibnamefont
  {{Fisher}}}, \bibinfo {author} {\bibfnamefont {P.~A.}\ \bibnamefont {{Lee}}},
  \bibinfo {author} {\bibfnamefont {N.}~\bibnamefont {{Nagaosa}}}, \ and\
  \bibinfo {author} {\bibfnamefont {X.-G.}\ \bibnamefont {{Wen}}},\ }\bibfield
  {title} {\enquote {\bibinfo {title} {{Stability of U(1) spin liquids in two
  dimensions}},}\ }\href {\doibase 10.1103/PhysRevB.70.214437} {\bibfield
  {journal} {\bibinfo  {journal} {Phys. Rev. B}\ }\textbf {\bibinfo {volume}
  {70}},\ \bibinfo {eid} {214437} (\bibinfo {year} {2004})},\ \Eprint
  {http://arxiv.org/abs/cond-mat/0404751} {cond-mat/0404751} \BibitemShut
  {NoStop}%
\bibitem [{\citenamefont {{Lee}}(2008)}]{SSLee08}%
  \BibitemOpen
  \bibfield  {author} {\bibinfo {author} {\bibfnamefont {S.-S.}\ \bibnamefont
  {{Lee}}},\ }\bibfield  {title} {\enquote {\bibinfo {title} {{Stability of the
  U(1) spin liquid with a spinon Fermi surface in 2+1 dimensions}},}\ }\href
  {\doibase 10.1103/PhysRevB.78.085129} {\bibfield  {journal} {\bibinfo
  {journal} {Phys. Rev. B}\ }\textbf {\bibinfo {volume} {78}},\ \bibinfo {eid}
  {085129} (\bibinfo {year} {2008})},\ \Eprint {http://arxiv.org/abs/0804.3800}
  {arXiv:0804.3800 [cond-mat.str-el]} \BibitemShut {NoStop}%
\bibitem [{\citenamefont {Metlitski}\ \emph {et~al.}(2015)\citenamefont
  {Metlitski}, \citenamefont {Mross}, \citenamefont {Sachdev},\ and\
  \citenamefont {Senthil}}]{MMSS14}%
  \BibitemOpen
  \bibfield  {author} {\bibinfo {author} {\bibfnamefont {M.~A.}\ \bibnamefont
  {Metlitski}}, \bibinfo {author} {\bibfnamefont {D.~F.}\ \bibnamefont
  {Mross}}, \bibinfo {author} {\bibfnamefont {S.}~\bibnamefont {Sachdev}}, \
  and\ \bibinfo {author} {\bibfnamefont {T.}~\bibnamefont {Senthil}},\
  }\bibfield  {title} {\enquote {\bibinfo {title} {{Cooper pairing in non-Fermi
  liquids}},}\ }\href {\doibase 10.1103/PhysRevB.91.115111} {\bibfield
  {journal} {\bibinfo  {journal} {Phys. Rev. B}\ }\textbf {\bibinfo {volume}
  {91}},\ \bibinfo {pages} {115111} (\bibinfo {year} {2015})},\ \Eprint
  {http://arxiv.org/abs/1403.3694} {arXiv:1403.3694 [cond-mat.str-el]}
  \BibitemShut {NoStop}%
\bibitem [{\citenamefont {{Zachar}}\ \emph {et~al.}(1998)\citenamefont
  {{Zachar}}, \citenamefont {{Kivelson}},\ and\ \citenamefont
  {{Emery}}}]{ZKE98}%
  \BibitemOpen
  \bibfield  {author} {\bibinfo {author} {\bibfnamefont {O.}~\bibnamefont
  {{Zachar}}}, \bibinfo {author} {\bibfnamefont {S.~A.}\ \bibnamefont
  {{Kivelson}}}, \ and\ \bibinfo {author} {\bibfnamefont {V.~J.}\ \bibnamefont
  {{Emery}}},\ }\bibfield  {title} {\enquote {\bibinfo {title} {{Landau theory
  of stripe phases in cuprates and nickelates}},}\ }\href {\doibase
  10.1103/PhysRevB.57.1422} {\bibfield  {journal} {\bibinfo  {journal} {Phys.
  Rev. B}\ }\textbf {\bibinfo {volume} {57}},\ \bibinfo {pages} {1422}
  (\bibinfo {year} {1998})},\ \Eprint {http://arxiv.org/abs/cond-mat/9702055}
  {cond-mat/9702055} \BibitemShut {NoStop}%
\bibitem [{\citenamefont {{Zaanen}}\ \emph {et~al.}(2001)\citenamefont
  {{Zaanen}}, \citenamefont {{Osman}}, \citenamefont {{Kruis}}, \citenamefont
  {{Nussinov}},\ and\ \citenamefont {{Tworzydlo}}}]{Zaanen01}%
  \BibitemOpen
  \bibfield  {author} {\bibinfo {author} {\bibfnamefont {J.}~\bibnamefont
  {{Zaanen}}}, \bibinfo {author} {\bibfnamefont {O.~Y.}\ \bibnamefont
  {{Osman}}}, \bibinfo {author} {\bibfnamefont {H.~V.}\ \bibnamefont
  {{Kruis}}}, \bibinfo {author} {\bibfnamefont {Z.}~\bibnamefont {{Nussinov}}},
  \ and\ \bibinfo {author} {\bibfnamefont {J.}~\bibnamefont {{Tworzydlo}}},\
  }\bibfield  {title} {\enquote {\bibinfo {title} {{The geometric order of
  stripes and Luttinger liquids}},}\ }\href {\doibase
  10.1080/13642810108208566} {\bibfield  {journal} {\bibinfo  {journal} {Phil.
  Mag. B}\ }\textbf {\bibinfo {volume} {81}},\ \bibinfo {pages} {1485}
  (\bibinfo {year} {2001})},\ \Eprint {http://arxiv.org/abs/cond-mat/0102103}
  {cond-mat/0102103} \BibitemShut {NoStop}%
\bibitem [{\citenamefont {{Nussinov}}\ and\ \citenamefont
  {{Zaanen}}(2002)}]{Zaanen02}%
  \BibitemOpen
  \bibfield  {author} {\bibinfo {author} {\bibfnamefont {Z.}~\bibnamefont
  {{Nussinov}}}\ and\ \bibinfo {author} {\bibfnamefont {J.}~\bibnamefont
  {{Zaanen}}},\ }\bibfield  {title} {\enquote {\bibinfo {title} {{Stripe
  fractionalization I: the generation of Ising local symmetry}},}\ }\href
  {\doibase 10.1051/jp4:20020405} {\bibfield  {journal} {\bibinfo  {journal}
  {J. Phys. IV France}\ }\textbf {\bibinfo {volume} {12}},\ \bibinfo {pages}
  {Pr9} (\bibinfo {year} {2002})},\ \Eprint
  {http://arxiv.org/abs/cond-mat/0209437} {cond-mat/0209437} \BibitemShut
  {NoStop}%
\bibitem [{\citenamefont {{Sachdev}}\ and\ \citenamefont
  {{Morinari}}(2002)}]{SSTM02}%
  \BibitemOpen
  \bibfield  {author} {\bibinfo {author} {\bibfnamefont {S.}~\bibnamefont
  {{Sachdev}}}\ and\ \bibinfo {author} {\bibfnamefont {T.}~\bibnamefont
  {{Morinari}}},\ }\bibfield  {title} {\enquote {\bibinfo {title} {{Strongly
  coupled quantum criticality with a Fermi surface in two dimensions:
  Fractionalization of spin and charge collective modes}},}\ }\href {\doibase
  10.1103/PhysRevB.66.235117} {\bibfield  {journal} {\bibinfo  {journal} {Phys.
  Rev. B}\ }\textbf {\bibinfo {volume} {66}},\ \bibinfo {eid} {235117}
  (\bibinfo {year} {2002})},\ \Eprint {http://arxiv.org/abs/cond-mat/0207167}
  {cond-mat/0207167} \BibitemShut {NoStop}%
\bibitem [{\citenamefont {{Mross}}\ and\ \citenamefont
  {{Senthil}}(2012{\natexlab{a}})}]{Mross12}%
  \BibitemOpen
  \bibfield  {author} {\bibinfo {author} {\bibfnamefont {D.~F.}\ \bibnamefont
  {{Mross}}}\ and\ \bibinfo {author} {\bibfnamefont {T.}~\bibnamefont
  {{Senthil}}},\ }\bibfield  {title} {\enquote {\bibinfo {title} {{Theory of a
  Continuous Stripe Melting Transition in a Two-Dimensional Metal: A Possible
  Application to Cuprate Superconductors}},}\ }\href {\doibase
  10.1103/PhysRevLett.108.267001} {\bibfield  {journal} {\bibinfo  {journal}
  {Phys. Rev. Lett.}\ }\textbf {\bibinfo {volume} {108}},\ \bibinfo {eid}
  {267001} (\bibinfo {year} {2012}{\natexlab{a}})},\ \Eprint
  {http://arxiv.org/abs/1201.3358} {arXiv:1201.3358 [cond-mat.str-el]}
  \BibitemShut {NoStop}%
\bibitem [{\citenamefont {{Mross}}\ and\ \citenamefont
  {{Senthil}}(2012{\natexlab{b}})}]{Mross12a}%
  \BibitemOpen
  \bibfield  {author} {\bibinfo {author} {\bibfnamefont {D.~F.}\ \bibnamefont
  {{Mross}}}\ and\ \bibinfo {author} {\bibfnamefont {T.}~\bibnamefont
  {{Senthil}}},\ }\bibfield  {title} {\enquote {\bibinfo {title} {{Stripe
  melting and quantum criticality in correlated metals}},}\ }\href {\doibase
  10.1103/PhysRevB.86.115138} {\bibfield  {journal} {\bibinfo  {journal} {Phys.
  Rev. B}\ }\textbf {\bibinfo {volume} {86}},\ \bibinfo {eid} {115138}
  (\bibinfo {year} {2012}{\natexlab{b}})},\ \Eprint
  {http://arxiv.org/abs/1207.1442} {arXiv:1207.1442 [cond-mat.str-el]}
  \BibitemShut {NoStop}%
\bibitem [{\citenamefont {{Kaul}}\ \emph
  {et~al.}(2008{\natexlab{b}})\citenamefont {{Kaul}}, \citenamefont
  {{Metlitski}}, \citenamefont {{Sachdev}},\ and\ \citenamefont
  {{Xu}}}]{RKK08b}%
  \BibitemOpen
  \bibfield  {author} {\bibinfo {author} {\bibfnamefont {R.~K.}\ \bibnamefont
  {{Kaul}}}, \bibinfo {author} {\bibfnamefont {M.~A.}\ \bibnamefont
  {{Metlitski}}}, \bibinfo {author} {\bibfnamefont {S.}~\bibnamefont
  {{Sachdev}}}, \ and\ \bibinfo {author} {\bibfnamefont {C.}~\bibnamefont
  {{Xu}}},\ }\bibfield  {title} {\enquote {\bibinfo {title} {{Destruction of
  N{\'e}el order in the cuprates by electron doping}},}\ }\href {\doibase
  10.1103/PhysRevB.78.045110} {\bibfield  {journal} {\bibinfo  {journal} {Phys.
  Rev. B}\ }\textbf {\bibinfo {volume} {78}},\ \bibinfo {eid} {045110}
  (\bibinfo {year} {2008}{\natexlab{b}})},\ \Eprint
  {http://arxiv.org/abs/0804.1794} {arXiv:0804.1794 [cond-mat.str-el]}
  \BibitemShut {NoStop}%
\bibitem [{\citenamefont {Whitsitt}\ and\ \citenamefont
  {Sachdev}(2016)}]{SWSS16}%
  \BibitemOpen
  \bibfield  {author} {\bibinfo {author} {\bibfnamefont {S.}~\bibnamefont
  {Whitsitt}}\ and\ \bibinfo {author} {\bibfnamefont {S.}~\bibnamefont
  {Sachdev}},\ }\bibfield  {title} {\enquote {\bibinfo {title} {Transition from
  the $\mathbb{Z}_{2}$ spin liquid to antiferromagnetic order: Spectrum on the
  torus},}\ }\href {\doibase 10.1103/PhysRevB.94.085134} {\bibfield  {journal}
  {\bibinfo  {journal} {Phys. Rev. B}\ }\textbf {\bibinfo {volume} {94}},\
  \bibinfo {pages} {085134} (\bibinfo {year} {2016})},\ \Eprint
  {http://arxiv.org/abs/1603.05652} {arXiv:1603.05652 [cond-mat.str-el]}
  \BibitemShut {NoStop}%
\bibitem [{\citenamefont {{Xu}}\ and\ \citenamefont
  {{Sachdev}}(2009)}]{CXSS09}%
  \BibitemOpen
  \bibfield  {author} {\bibinfo {author} {\bibfnamefont {C.}~\bibnamefont
  {{Xu}}}\ and\ \bibinfo {author} {\bibfnamefont {S.}~\bibnamefont
  {{Sachdev}}},\ }\bibfield  {title} {\enquote {\bibinfo {title} {{Global phase
  diagrams of frustrated quantum antiferromagnets in two dimensions: Doubled
  Chern-Simons theory}},}\ }\href {\doibase 10.1103/PhysRevB.79.064405}
  {\bibfield  {journal} {\bibinfo  {journal} {Phys. Rev. B}\ }\textbf {\bibinfo
  {volume} {79}},\ \bibinfo {eid} {064405} (\bibinfo {year} {2009})},\ \Eprint
  {http://arxiv.org/abs/0811.1220} {arXiv:0811.1220 [cond-mat.str-el]}
  \BibitemShut {NoStop}%
\bibitem [{\citenamefont {Lehoucq}\ \emph {et~al.}(2003)\citenamefont
  {Lehoucq}, \citenamefont {Uzan},\ and\ \citenamefont {Weeks}}]{lehoucq2003}%
  \BibitemOpen
  \bibfield  {author} {\bibinfo {author} {\bibfnamefont {R.}~\bibnamefont
  {Lehoucq}}, \bibinfo {author} {\bibfnamefont {J.-P.}\ \bibnamefont {Uzan}}, \
  and\ \bibinfo {author} {\bibfnamefont {J.}~\bibnamefont {Weeks}},\ }\bibfield
   {title} {\enquote {\bibinfo {title} {Eigenmodes of lens and prism spaces},}\
  }\href {\doibase 10.2996/kmj/1050496653} {\bibfield  {journal} {\bibinfo
  {journal} {Kodai Math. J.}\ }\textbf {\bibinfo {volume} {26}},\ \bibinfo
  {pages} {119} (\bibinfo {year} {2003})}\BibitemShut {NoStop}%
\end{thebibliography}%

\end{document}